\documentclass[aps,prx,reprint,10pt,twocolumn,citeautoscript,superscriptaddress]{revtex4-2}
\pdfoutput=1

\usepackage{natbib}
\usepackage[english]{babel}
\usepackage{letltxmacro}
\usepackage{latexsym}
\LetLtxMacro{\ORIGselectlanguage}{\selectlanguage}
\makeatletter
\DeclareRobustCommand{\selectlanguage}[1]{%
  \@ifundefined{alias@\string#1}
    {\ORIGselectlanguage{#1}}
    {\begingroup\edef\x{\endgroup
       \noexpand\ORIGselectlanguage{\@nameuse{alias@#1}}}\x}%
}
\newcommand{\definelanguagealias}[2]{%
  \@namedef{alias@#1}{#2}%
}
\makeatother
\definelanguagealias{en}{english}
\definelanguagealias{English}{english}
\usepackage{graphicx}
\usepackage{amsmath}
\usepackage{amsfonts}
\usepackage{amssymb}
\usepackage{bm}
\usepackage{color}
\usepackage[percent]{overpic}
\usepackage{soul} 
\usepackage{amssymb}
\usepackage{wasysym}
\usepackage{dsfont}
\usepackage{float}
\usepackage[mathscr]{euscript}
\usepackage{lipsum}
\usepackage{hyperref}
\hypersetup{
    pdfstartview={FitH},    
    colorlinks=true,       
    linkcolor=blue,          
    citecolor=blue,        
    filecolor=magenta,      
    urlcolor=blue           
}

\newcommand{\be}{\begin{equation}}
\newcommand{\ee}{\end{equation}}
\newcommand{\bea}{\begin{eqnarray}}
\newcommand{\eea}{\end{eqnarray}}

\newcommand{\mc}{\mathcal}

\newcommand{\vect}[1]{\boldsymbol{#1}}

\usepackage{graphicx}
\usepackage[colorinlistoftodos]{todonotes}
\usepackage{verbatim}
\usepackage[normalem]{ulem}
\usepackage{enumitem}

\begin{document}

\title{Anomalous suppression of large-scale density fluctuations in classical and quantum spin liquids}

\author{Duyu Chen}
\thanks{These authors contributed equally to this work.\\
$^\dagger$Email:
\href{mailto:torquato@electron.princeton.edu}{torquato@electron.princeton.edu}}
\affiliation{Materials Research Laboratory, University of California, Santa
Barbara, CA 93106}

\author{Rhine Samajdar}
\thanks{These authors contributed equally to this work.\\
$^\dagger$Email:
\href{mailto:torquato@electron.princeton.edu}{torquato@electron.princeton.edu}}
\affiliation{Department of Physics, Princeton University, Princeton, NJ 08544}
\affiliation{Princeton Center for Theoretical Science, Princeton University, Princeton, NJ 08544}

\author{Yang Jiao}
\affiliation{Materials Science and Engineering, Arizona State University,
Tempe, AZ 85287}
\affiliation{Department of Physics, Arizona State University, Tempe, AZ 85287}

\author{Salvatore Torquato$^\dagger$}
\affiliation{Department of Physics, Princeton University, Princeton, NJ 08544}
\affiliation{Department of Chemistry, Princeton University, Princeton,
NJ 08544}
\affiliation{Princeton Materials Institute, Princeton University, Princeton, NJ 08540}
\affiliation{Program in Applied and Computational Mathematics, Princeton, NJ 08544}

\date{\today}

\begin{abstract}

Classical spin liquids (CSLs) are intriguing states of matter that do not exhibit long-range magnetic order and are characterized by an extensive ground-state degeneracy. Adding quantum fluctuations, which induce dynamics between these different classical ground states, can give rise to quantum spin liquids (QSLs). QSLs are highly entangled quantum phases of matter characterized by fascinating emergent properties, such as fractionalized excitations and topological order. One such exotic quantum liquid is the $\mathbb{Z}_2$ QSL, which can be regarded as a resonating valence bond (RVB) state formed from superpositions of dimer coverings of an underlying lattice. In this work, we unveil a \textit{hidden} large-scale structural property of archetypal CSLs and QSLs known as hyperuniformity, i.e., normalized infinite-wavelength density fluctuations are completely suppressed in these systems. In particular, we first demonstrate that classical ensembles of close-packed dimers and their corresponding quantum RVB states are perfectly hyperuniform in general. Subsequently, we focus on a ruby-lattice spin liquid that was recently realized in a Rydberg-atom quantum simulator, and show that the QSL remains effectively hyperuniform even in the presence of a finite density of spinon and vison excitations, as long as the dimer constraint is still largely preserved. Moreover, we demonstrate that metrics based on the framework of hyperuniformity can be used to distinguish the QSL from other proximate quantum phases. These metrics can help identify potential QSL candidates, which can then be further analyzed using more advanced, computationally-intensive quantum numerics to confirm their status as true QSLs.

\end{abstract}

\maketitle

The notion of hyperuniformity, introduced two decades ago \cite{To03}, provides a powerful, unified framework to classify and characterize different ordered systems (crystals and quasicrystals) as well as unusual disordered ones \cite{To03, To18a}. The remarkable feature of hyperuniform many-body systems is the complete suppression of (normalized) density fluctuations at large length scales. Specifically, this means that in $d$-dimensional Euclidean space 
$\mathbb{R}^d$, the normalized local number variance $\sigma^2(R)/R^d$\,$\rightarrow$\,$0$ in the large-$R$ limit, where $\sigma^2(R)$\,$\equiv$\,$\langle N^2(R)\rangle$\,$-$\,$\langle N(R) \rangle^2$ is the local number variance, $N(R)$ represents the number of particles in a spherical window of radius $R$ randomly placed into the system, and $\langle \cdots \rangle$ denotes an ensemble average \cite{To03, To18a}. Equivalently, the static structure factor $S({\boldsymbol{k}})$---which is proportional to the scattering intensity in X-ray-, light-, or neutron-scattering experiments (excluding forward scattering)---vanishes in the infinite-wavelength limit for hyperuniform systems, i.e., $\lim_{k\rightarrow 0}S(k) = 0$, where the wavenumber $k=|{\boldsymbol{k}}|$ is the magnitude of the wavevector ${\boldsymbol{k}}$. In fact, the small-$k$ scaling behavior of $S(k) \sim k^\alpha$ ($\alpha > 0$ for hyperuniform systems) dictates the large-$R$ asymptotic behavior of $\sigma^2(R)$, based on which all hyperuniform systems can be categorized into three classes: $\sigma^2(R) \sim R^{d-1}$ for $\alpha>1$ (class I); $\sigma^2(R) \sim R^{d-1}\ln R$ for $\alpha=1$ (class II); and $\sigma^2(R) \sim R^{d-\alpha}$ for $0<\alpha<1$ (class III) \cite{To18a}. Typical disordered systems, such as ordinary liquids and glasses, on the other hand, are not hyperuniform, and possess a large-$R$ scaling of $\sigma^2(R) \sim R^d$ \cite{To18a}. 

Disordered hyperuniform systems are exotic states of matter \cite{To03, To18a} that lie between perfect crystals and typical liquids. These unusual amorphous systems are similar to liquids or glasses in that they are statistically isotropic and possess no Bragg peaks, and yet they completely suppress normalized large-scale density fluctuations, like crystals, and in this sense, possess a hidden long-range order \cite{To03, Za09, To18a}. Disordered hyperuniform states have been discovered in a variety of equilibrium and nonequilibrium classical physical and biological systems, and appear to endow such systems \cite{To03, To18a, Ji14, Dr15, He15, Ja15, Ru19, Le19a, Le19b, Hu21, Li24} with desirable optical, transport, and mechanical properties that cannot be achieved in either ordinary disordered or perfectly crystalline states \cite{To03, To18a, Fl09, Ma13, Le16, Zh16, Ch18a, Xu17, Ki20, Ki23, Kl22, Au20, Ro21, Ae22, Ch22, Ta22}. While the hyperuniformity of classical systems has been extensively investigated, its potential extension to quantum systems has been much less explored. Besides families of determinantal point processes \cite{To08,Sc09,Ab17,Ma21} (which can exactly attain the $n$-point correlation functions of certain quantum ground states), to date, there are only a handful of quantum materials and models \cite{Fe56,Ge19,Ll20,Sa22a,Sa22b,Wa24} that are known to exhibit hyperuniformity. Here, we make a distinction between such emergent hyperuniformity and quantum systems that are explicitly engineered to be hyperuniform by an appropriate choice of couplings \cite{Cr19,Bo21}. 

A classical spin liquid (CSL) is a system characterized by an extensive ground-state degeneracy and the absence of long-range magnetic order \cite{Ya23}. In recent years, CSLs have attracted much interest, since they are closely connected to the finite-temperature physics of frustrated magnets \cite{Ya20, Be21, Ya23}. A quantum spin liquid (QSL), on the other hand, arises when quantum fluctuations introduce dynamics between the classical ground states of a CSL \cite{Ya23}. The long-range many-body quantum entanglement structure inherent in a QSL endows it with a host of intriguing emergent properties, ranging from fractionalized excitations to topological order \cite{savary2016quantum,knolle2019field,broholm2020quantum}. The simplest possible QSL that preserves all symmetries, including time reversal, is the $\mathbb{Z}_2$ quantum spin liquid \cite{ReadSachdev91,Wen91,Sachdev92}, which is a gapped state with the same topological order as the celebrated toric code \cite{kitaev2006anyons}. Many aspects of CSLs and QSLs have been extensively studied over the last few decades \cite{Ya20, Be21, Ya23, savary2016quantum,knolle2019field,broholm2020quantum}, but a common feature of these systems is that they usually possess no apparent long-range spatial order. While the concept of hyperuniformity has been hitherto unexplored in the context of these unordered states of matter, it could potentially shed light on the fundamental understanding of the nature of long-range correlations in these systems, and provide a framework to distinguish different disordered and ordered phases.

Despite decades of efforts to find $\mathbb{Z}_2$ QSLs in solid-state materials \cite{lee2008end}, their direct experimental detection has largely been elusive. Recently, such a $\mathbb{Z}_2$ QSL was realized using a programmable quantum simulator based on arrays  of neutral atoms coupled to highly excited Rydberg states \cite{Semeghini.2021}. Due to strong interatomic dipole-dipole interactions, in this system, the excitation of one atom to the Rydberg state prohibits the simultaneous excitation of other atoms located within a so-called ``blockade radius'' \cite{jaksch2000fast}. This robust phenomenon, known as the ``Rydberg blockade'', naturally gives rise to strong correlations between the atoms, leading to a variety of interesting quantum phases \cite{bernien2017probing,de2019observation,Samajdar_2020,PhysRevLett.126.170603,Ebadi.2021,scholl2021quantum,Kim.2021}, phase transitions \cite{samajdar2018numerical,whitsitt2018quantum,PhysRevLett.122.017205,chepiga2021kibble,kalinowski2021bulk}, and many-body dynamics \cite{turner2018weak,keesling2019quantum,Bluvstein.2021}.

\begin{figure}[tb]
\includegraphics[width=\linewidth]{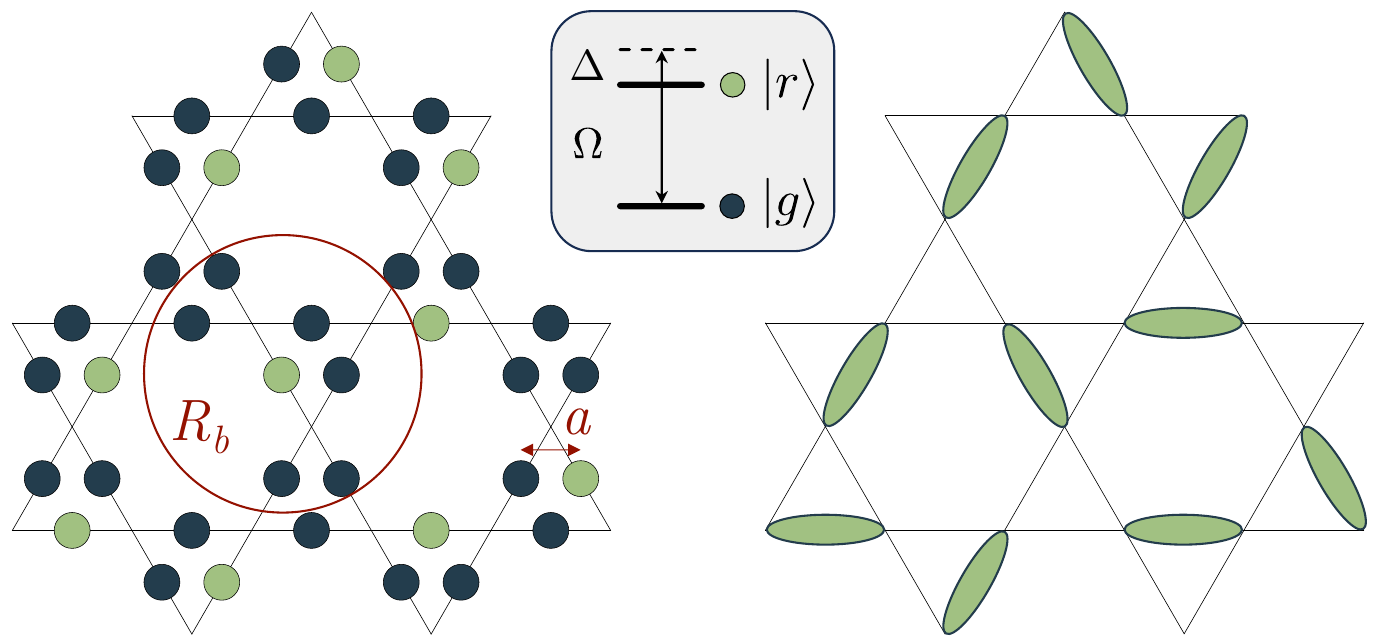}
\caption{Mapping between a configuration of Rydberg atoms on the ruby lattice (left) and a dimer covering of the kagome lattice (right). The blockade radius $R_b$ is chosen so as to encompass two nearest neighbors, two second-nearest neighbors, and two third-nearest neighbors of an atom in the excited state $\rvert r\rangle$ (on the ruby lattice), which implements a constraint of having exactly one dimer touching each vertex of the kagome lattice.}
\label{fig:1}
\end{figure}

The realization of a Rydberg $\mathbb{Z}_2$ QSL is made possible by a mapping \cite{Samajdar.2021,Verresen.2020,yan2022triangular} between Rydberg atoms and quantum dimer models (QDMs), which are parent Hamiltonians known to host exact QSL ground states \cite{moessner2001ising,moessner2008quantum}. In particular, Fig.~\ref{fig:1} illustrates a mapping between the excitations of Rydberg atoms positioned on a ruby lattice and quantum dimers  on the bonds of the kagome lattice. Note that the ruby lattice is the medial lattice (or line graph) of the kagome lattice, i.e., the sites of the former are the centers of the bonds of the latter \cite{misguich2002quantum}. The coherent quantum superposition of all possible blockade-allowed configurations of Rydberg excitations on the ruby lattice is thus equivalent to a ``resonating valence bond'' (RVB) liquid state described by a macroscopic superposition of dimer coverings of the kagome lattice with exactly one dimer touching each vertex \cite{misguich2002quantum,misguich2003quantum,misguich2004interaction}, which \textit{is} the aforementioned $\mathbb{Z}_2$ QSL.

In this work, we use the formalism of  hyperuniformity to probe the structural properties of representative CSLs and QSLs and quantitatively unveil the unique nature of the ``liquidity'' of these systems. We show that although CSLs and QSLs lack symmetry-breaking spatial order, akin to a conventional \textit{liquid}, they are also characterized by an anomalous suppression of long-wavelength density fluctuations, as expected for a \textit{crystal}. In particular, in Sec.~\ref{sec2}, we demonstrate that a classical ensemble of kagome-lattice dimer coverings is perfectly hyperuniform, with a structure factor $S(k)$ that scales as $S(k) \sim k^6$ at small $k$. In Sec.~\ref{sec3}, we first show that fixed-point RVB states, the quantum counterparts of classical dimer coverings, are also perfectly hyperuniform in general. Then, we illustrate that the QSL remains effectively hyperuniform (i.e., hyperuniform in the practical sense according to the metric $B/A$, defined in Sec.~\ref{sec3}), upon introducing a finite density of quasiparticle excitations (namely, ``spinons'', or monomers) into the perfect RVB state, which are inevitable in any realistic setting. Our results thus demonstrate how notions of hyperuniformity can be used to distinguish between QSLs and proximate disordered but nontopological phases in a quantum phase diagram. Besides strengthening our fundamental understanding of density fluctuations in CSLs and QSLs, our work also strikingly demonstrates hyperuniformity for a strongly interacting, long-range entangled quantum system.

\section{Classical spin liquids}\label{sec2}
An archetypal example of a classical spin liquid is the set of all dimer coverings of a given lattice subject to a hard-core constraint, i.e., each site participates in a dimer with one, and only one, of its nearest neighbors. While intriguing on their own right---as discussed above---such CSLs also form the building blocks for quantum RVB states. Here, we start by considering the ensemble of all possible dimer coverings of the kagome lattice, which, as mentioned previously, are in one-to-one correspondence with the allowed arrangements of Rydberg atoms on the ruby lattice when up to third-nearest-neighboring sites are blockaded. There are $\mathcal{D} \equiv 2^{N/3+1}$ such coverings \cite{elser1993kagome,misguich2003quantum} (where $N$ is the number of ruby-lattice sites), each of which represents a degenerate ground state of a classical fully frustrated Ising model on the dice lattice \cite{fisher1966dimer,mossner2003ising}. Owing to this extensive ground state degeneracy, which leads to a nonzero entropy even at zero temperature, the statistical-mechanical ensemble of such dimer coverings can be regarded as a classical spin liquid \cite{balents2010spin}.

\begin{figure}[tb]
\includegraphics[width=\linewidth]{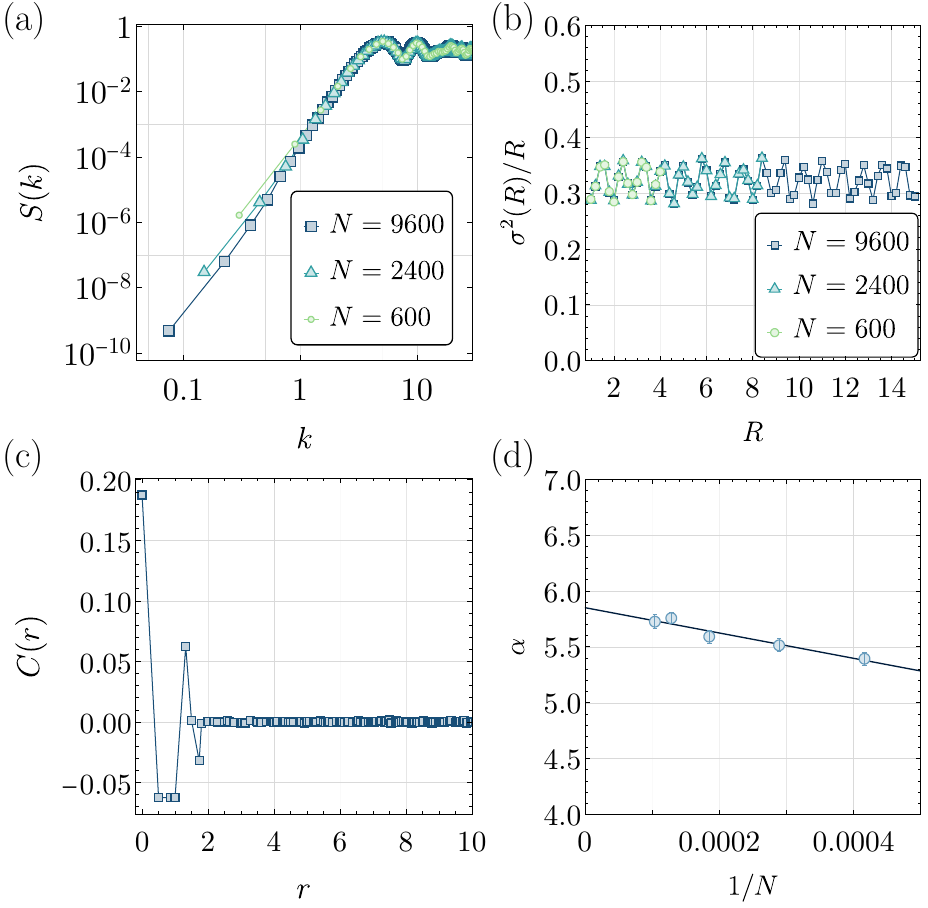} 
\caption{(a) Structure factor $S(k)$ of perfect classical dimer coverings on the kagome lattice at different system sizes $N$. (b) Local number variance $\sigma^2(R)$ of perfect dimer coverings on the kagome lattice at different system sizes $N$. (c) Dimer-dimer pair correlation function $C(r)$ of perfect dimer coverings on the kagome lattice at $N = 9600$. (d) Small-$k$ scaling exponent $\alpha$ of $S(k)$ as a function of system size $N$ for perfect dimer coverings on the kagome lattice. In all these plots, we work in units where the side length of a hexagon in the kagome lattice is set to unity.} 
\label{fig_2}
\end{figure}

Focusing on the kagome lattice, we first enforce the Rydberg blockade constraint by imposing a hard-core repulsion between dimers that allows each kagome vertex to be covered by at most one dimer. Then, the maximum filling fraction of $1/4$ (on the ruby lattice) corresponds to a ``perfect'' (monomer-free) dimer covering of the kagome lattice. We develop a simulated annealing algorithm (see Methods for details) to generate such perfect dimer coverings at $f$\,$=$\,$1/4$ for various system sizes $N = 600$, 2400, 3456, 5400, 7776, and 9600 (see Fig. \ref{fig:fig_6} for a representative configuration at $N = 2400$), and compute their pair statistics. Here, we employ periodic boundary conditions in both the horizontal and vertical directions. Fig.~\ref{fig_2}(a) shows the ensemble-averaged structure factor $S(k)$ of the dimer centers (corresponding to Rydberg excitations in the atomic picture) for a few representative system sizes ($N$\,$=$\,$600$, 2400, and $9600$), which clearly approaches zero as $k$ goes to zero, indicating perfect hyperuniformity of the perfect dimer coverings. Our work reports the discovery of a disordered hyperuniform ``packing'' on a lattice subject to well-defined packing constraints, which is different from previous reports of disordered hyperuniform structures on lattices (obtained, e.g., via inverse design techniques) \cite{Di18a,Ch18a,Ch23}. It is also noteworthy that, by definition, $S(k)$ is the same for unoccupied edges (or atoms in the ground state) and occupied edges (or atoms in the Rydberg state). Consequently, the spatial distribution of unoccupied edges is perfectly hyperuniform as well.

In Fig.~\ref{fig_2}(b), we present the corresponding ensemble-averaged local number variance $\sigma^2(R)$, which scales linearly as $R$ increases and further demonstrates that perfect dimer coverings are class-I hyperuniform. We also compute the dimer-dimer pair correlation function $C(r)$ defined as $C(r)\equiv \langle n(x)n(x+r) \rangle - \langle n(x)\rangle^2$ for the system of size $N$\,$=$\,$9600$, with the result sketched in Fig.~\ref{fig_2}(c). Here, $n(x)$ denotes the occupation at position $x$, which equals 1 if the site is occupied by a dimer and 0 otherwise, and $\langle \cdots \rangle$ denotes an ensemble and volume average over a single ensemble member. We implicitly assume that we can perform an angular average of $C({\vect{r}})$ to obtain $C(r)$, where ${\vect{r}}$ is the vector displacement and $r=|{\vect{r}}|$ is the magnitude of ${\vect{r}}$; in the Supplemental Information (SI), we show why such an angular average is physically meaningful. Interestingly, $C(r)$ decays to its long-range value of zero very rapidly as $r$ increases. This is consistent with prior theoretical results on the corresponding quantum dimer liquid, for which dimer-dimer correlations are known to be exactly zero for a pair of dimers when their corresponding triangles do not share a common vertex \cite{misguich2002quantum}. We note that an exponential or faster decay of $C(r)$, as seen here, means that the structure factor $S(k)$ is analytic at the origin and hence, the small-$k$ scaling exponent $\alpha$ of $S(k)$ should be a non-negative even integer \cite{To18a}. Once again, we make the implicit assumption that we can perform an angular average of $S({\vect{k}})$ to obtain $S(k)$, as justified in the SI. Computing the exponent $\alpha$ from the large-$t$ scaling behavior of the excess spreadability $\mathcal{S}(\infty) - \mathcal{S}(t)$ (see SI for details) \cite{Wa22} for various system sizes $N$ [Fig.~\ref{fig_2}(d)], and numerically extrapolating to the infinite-system limit, we find that $\alpha \approx 6$ as $N\rightarrow \infty$; combining this observation with the analyticity of $S(k)$ at $k=0$ allows us to conclude that $\alpha = 6$ for perfect dimer coverings on the kagome lattice in the infinite-system limit. Furthermore, in the SI, we also theoretically prove that $\alpha=6$ in this case.

We now briefly outline an intuitive physical argument for \textit{why} the ensemble of perfect dimer coverings on the kagome lattice should be perfectly hyperuniform and disordered. At $f = 1/4$, each vertex of the kagome lattice is touched by exactly one dimer, and accordingly, each dimer covers exactly two vertices of the kagome lattice. If one randomly throws circular observation windows of radius $R$ into the system, and counts the number of kagome vertices as well as the number of dimers that fall within each window, the former is exactly two times the latter within the window, except for fluctuations concentrated near the surface of the window, implying a local number variance $\sigma^2(R)$ that scales like $R$ for large $R$ and hence class-I hyperuniformity (see SI for a detailed rigorous theoretical analysis). This also means that the small-$k$ scaling exponent $\alpha$ of $S(k)$ must be larger than 1 \cite{To18a}. Why are such perfect  hyperuniform dimer coverings also disordered? The answer is because of the infinite number of arrangements of dimer coverings (high entropy configurations)  consistent with local blockade constraints in the infinite-volume limit.

\section{Quantum spin liquids}\label{sec3}

\subsection{Quantum resonating valence bond states}
So far, our analysis has examined a classical ensemble of kagome-lattice dimer coverings, without any notion of phase coherence between individual configurations.
Quantum fluctuations introduce dynamics between these exponentially many classical ground states. In order to define a \textit{quantum} spin liquid, we first promote the set of allowed classical dimer configurations to quantum states $\rvert q_\alpha\rangle$, which constitute the orthonormal basis vectors of a Hilbert space. The quantum nature is naturally reflected in that this vector space now also allows states that are linear superpositions of dimer coverings. 

An example of a quantum liquid, which preserves all spatial symmetries, is the celebrated resonating valence bond (RVB) state \cite{pauling1949resonating,anderson1973resonating,baskaran1988gauge,rokhsar1988superconductivity}. An ansatz for such an RVB state can be written as $\lvert \Psi_{\textsc{rvb}}\rangle$\,$=$\,$\sum_{\alpha=1}^{\mc{D}}c_\alpha \lvert q_\alpha\rangle$, where $c_\alpha$\,$\in$\,$\mathbb{C}$ and $\mc{D}$ is the number of quantum states in the macroscopic superposition. We assume that the coefficients $c_\alpha$ respect the translational and rotational symmetries of the underlying lattice. We also note that we define $\lvert q_\alpha\rangle \equiv \lvert n_1 n_2 \cdots n_N \rangle$ (subject to the packing constraint of dimers), where $n_i = 1$ if site $i$ is occupied by a dimer and $n_i = 0$ otherwise. This definition ensures that different dimer coverings $\lvert q_\alpha\rangle$ are orthogonal \cite{Se13}, by construction, as opposed to the case where the RVB state is expressed as the superposition of different singlet pair states that are not orthogonal \cite{Su88, misguich2003quantum}. A theory for a stable, gapped RVB state with time-reversal symmetry was originally developed in Refs.~\cite{ReadSachdev91,Wen91,Sachdev92}, which first described the $\mathbb{Z}_2$ spin liquid.

An extended $\mathbb{Z}_2$ QSL phase was discovered in the kagome-lattice quantum dimer model \cite{misguich2002quantum,misguich2003quantum,misguich2004interaction}.
In this QSL phase, the ground-state wavefunction is generically of the form of $\lvert \Psi_{\textsc{rvb}}\rangle$ but the coefficients $c_\alpha$ may vary across parameter space. However, as long as each quantum state $\lvert q_\alpha\rangle$ is a perfect dimer covering at the maximal filling fraction, the resulting RVB state should also be perfectly hyperuniform (since each individual covering is perfectly hyperuniform per our theoretical analysis in the SI). In other words, the hyperuniformity of the QSL is not destroyed by the presence of flux-like ``vison'' \cite{read1989statistics,SenthilFisher} excitations, which are marked by phase flips between different dimer configurations.

\begin{figure}[tb]
\includegraphics[width=1.0\linewidth]{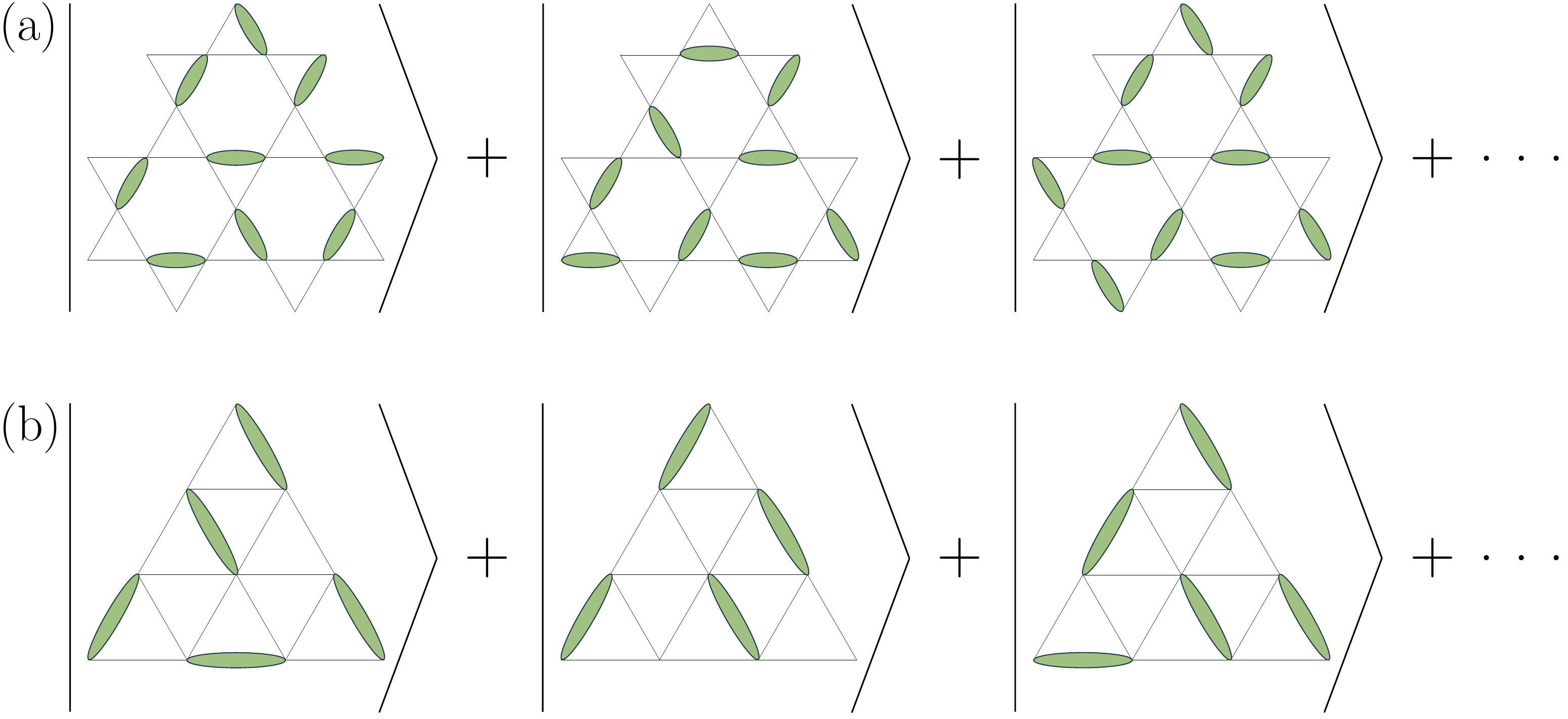} 
\caption{Illustration of fixed-point quantum spin liquids, constructed by the quantum superposition of perfect dimer coverings in the same topological sector, on the (a) kagome and (b) triangular lattices.} \label{fig_3}
\end{figure}

A special case of the aforementioned RVB state is the so-called Rokhsar-Kivelson (RK) point of the QDM Hamiltonian. At this point, the model is exactly solvable and the ground state is defined by an equal-amplitude superposition of all dimer coverings in a topological sector \cite{sutherland1988systems,misguich2002quantum}, i.e., $\lvert \Psi \rangle = \sum_{\alpha=1}^{\mc{D}_s} \lvert q_\alpha\rangle/\sqrt{\mc{D}_s}$, where $\mc{D}_s$ is the number of states in a sector $s$. In Fig.~\ref{fig_3}(a), we illustrate such a QSL state with dimers living on the kagome lattice. 
Now, it is easy to observe that when $\lvert c_1 \rvert = \lvert c_2 \rvert=\cdots=\lvert c_{\mc{D}_s}\rvert =1/\sqrt{\mc{D}_s}$, the structure factor $S_{\mathrm{q}}(k)$ of the quantum spin liquid is given by
\begin{equation}
	S^{}_{\mathrm{q}}(k) = \sum_{\alpha=1}^{\mc{D}_s}\lvert c^{}_\alpha \rvert^2 S^{}_\alpha(k) = \frac{1}{\mc{D}_s} \sum_{\alpha=1}^{\mc{D}_s} S^{}_\alpha(k) = \left \langle S^{}_{\mathrm{c}}(k) \right\rangle^{}_s,
\end{equation}
where $\langle S_{\mathrm{c}}(k)\rangle_s$ is the ensemble-averaged structure factor of the classical set of perfect dimer coverings belonging to the topological sector $s$. However, since the pair statistics $C(r)$ and $S(k)$ are the same for classical dimer coverings in different sectors, we can drop the label $s$ and obtain the ensemble-averaged structure factor by averaging over the full set of perfect dimer coverings, without differentiating between topological sectors $s$; as a result $S_{\mathrm{q}}(k) = \langle S_{\mathrm{c}}(k) \rangle$. Therefore, the previously established  perfect hyperuniformity of the classical spin liquid translates to the perfect hyperuniformity of this ``fixed-point'' quantum spin liquid at the RK point of the kagome-lattice QDM.

Similarly, we can construct RK QSLs on other lattices by considering superpositions of perfect dimer coverings on different lattices. For instance, in Fig.~\ref{fig_3}(b) we show a quantum spin liquid state with exactly one dimer touching each vertex of the triangular lattice \cite{moessner2001resonating}, which corresponds to a filling fraction of $1/6$ for Rydberg excitations on the kagome lattice \cite{Samajdar.2021}. Following the same theoretical arguments as those in Sec.~\ref{sec2}, these ideal QSL states on other lattices should also be class-I hyperuniform. 
However, the exact even-integer value of $\alpha$ may vary from lattice to lattice. For example, through rigorous finite-size-scaling analyses (see SI for details) we have determined the scaling exponent to be $\alpha= 4$ for perfect dimer liquids on the triangular lattice.

\subsection{Quantum fluctuations and matter fields}


In the preceding section, we demonstrated the hyperuniformity of a fixed-point $\mathbb{Z}_2$ QSL state, which is the ground state of a quantum dimer model and corresponds to the deconfined phase of a pure $\mathbb{Z}_2$ gauge theory \cite{MSF02}. However, a more realistic QSL (as might be realized in experiments) is described by a gauge theory coupled to matter fields, and the consequence of this distinction for hyperuniformity is an important question that requires scrutiny.

To this end, we consider the system of interacting Rydberg atoms on a ruby lattice, which was demonstrated to host a $\mathbb{Z}_2$ QSL phase in recent experiments \cite{Semeghini.2021}. Such an array is described by the spin-1/2 Hamiltonian, 
\begin{equation}
\label{eq:HRyd}
H = \frac{\Omega}{2} \sum_i \sigma^x_i - \Delta \sum_i n^{}_i  + \sum_{i<j} V^{}_{ij} n^{}_i n^{}_j,
\end{equation}
where $n_i \equiv (\sigma^z_i + 1)/2$, with $\sigma^z_i = +1$ ($-1$) corresponding to the presence (absence) of a Rydberg excitation on site $i$ of the ruby lattice.
Here, $\Omega$ denotes the Rabi frequency characterizing the coherent oscillations between the ground state $\rvert g\rangle$ and the excited state $\rvert r\rangle$, $\Delta$ is the laser detuning, and $V_{ij}$\,$=$\,$V_0/\lvert\vect{r}_i-\vect{r}_j\rvert^6$ is the van der Waals interaction between two excited atoms that are located at positions $\vect{r}_i, \vect{r}_j$. The strength of the interactions can be conveniently parameterized by the blockade radius, defined according to the relation $V_0 \equiv \Omega R_b^6$. Previously, it was found \cite{Semeghini.2021} that the realistic model incorporating the van der Waals interaction behaves similarly to a so-called PXP model, in which the hard-core exclusion between two excited atoms within the blockade radius is strictly enforced by an infinite energy penalty. Therefore, in this work, for simplicity, we choose to work with the PXP model, which yields the same physics.

In the following, we choose $R_b$ such that pairs of Rydberg atoms up to third-nearest neighbors are blockaded. The associated blockade constraint ensures that there can be \textit{at most} one dimer touching each vertex of the kagome lattice (unlike the hard constraint of the QDM, which asserts that there is \textit{exactly} one dimer touching each vertex of the kagome lattice). On the ruby lattice, our choice of $R_b$ enforces that the maximum filling fraction of Rydberg excitations is $1/4$. However, the number of excitations, $N_{\mathrm{exc}}$, is not conserved by the Hamiltonian \eqref{eq:HRyd}, so the actual filling fraction $\langle n_i \rangle \le 1/4$ varies depending on the precise point in parameter space. Therefore, at low energies, the Hilbert space of this model has to be expanded to allow for monomers, which are referred to as ``spinons''. In fact, the quantum-mechanical ground state necessarily has a small nonzero density of spinons \cite{PhysRevLett.130.043601}.

\begin{figure}[tb]
\includegraphics[width=\linewidth]{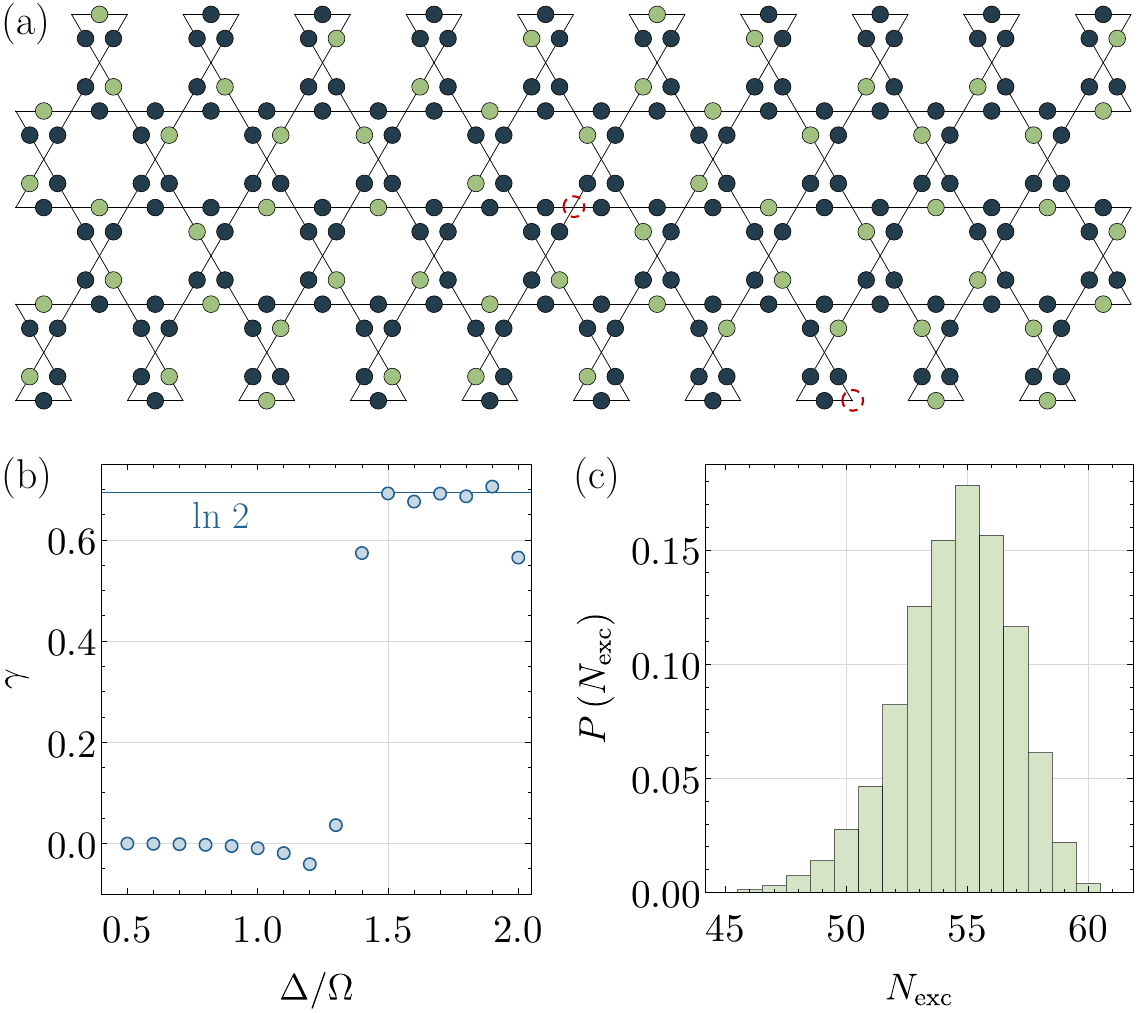} 
\caption{(a) A representative snapshot of Rydberg excitations in a system consisting of 240 sites at $\Delta/\Omega$ = 1.7; the monomers arising due to quantum fluctuations are circled in red. (b) Topological entanglement entropy $\gamma$ as a function of $\Delta/\Omega$. (c) The distribution $P(N_{\mathrm{exc}})$ of the number of excitations in a system consisting of 240 sites at $\Delta/\Omega$ = 1.7 obtained by sampling 10000 snapshots.} \label{fig_4}
\end{figure}

Employing the density-matrix renormalization group (DMRG) algorithm \cite{white1992density,white1993density,schollwock2005density,schollwock2011density}, we compute the system's ground-state wavefunction on long cylinders---with up to 288 atoms---as a function of $\Delta/\Omega$, keeping $R_b/a$ fixed as noted above. From the numerically obtained tensor-network representation of the ground state, we then generate 10,000 individual ``snapshots'' by sampling the wavefunction according to the Born rule; one such representative configuration for $N$\,$=$\,$240$ atoms is sketched in Fig.~\ref{fig_4}(a). Physically, the sampling is equivalent to performing a projective measurement in the $\sigma^{z}$ basis, which collapses the wavefunction to a particular monomer-dimer covering from among the exponentially many components constituting the ground-state wavefunction. This set of snapshots, which is analogous to the classical ensemble of dimer coverings studied in Sec.~\ref{sec2}, provides a natural starting point for studying the possible hyperuniformity of the Rydberg $\mathbb{Z}_2$ QSL phase. As $\Delta/\Omega$ varies, we observe that the Hamiltonian \eqref{eq:HRyd} hosts three distinct quantum phases \cite{Verresen.2020}: a trivial disordered phase [a quantum paramagnet that visually resembles a typical nonhyperuniform classical fluid \cite{To18a}, as we quantify below] for $\Delta/\Omega \lesssim 1.5$, the $\mathbb{Z}_2$ QSL for $1.5 \lesssim \Delta/\Omega \lesssim 2.0$, and an ordered ``valence bond solid'' (VBS) at large $\Delta/\Omega \gtrsim 2.0$ [for which only a few dimer configurations dominate in the ground state \cite{nikolic2003physics,singh2007ground,poilblanc2011competing}]. These three phases are also reflected in the topological entanglement entropy $\gamma$ \cite{PhysRevLett.96.110404}, shown in Fig.~\ref{fig_4}(b), which is theoretically known to be $\ln 2$ for a $\mathbb{Z}_2$ QSL \cite{jiang2012identifying} and zero for both the disordered and VBS phases. However, the system sizes accessible to today's state-of-the-art numerical methods, including DMRG and quantum Monte Carlo, are much smaller than that required to reliably ascertain hyperuniformity or lack thereof.

In order to overcome this limitation, we first recognize that the key feature of the realistic monomer-dimer $\mathbb{Z}_2$ QSL---vis-à-vis the fixed-point QDM liquid---is that the excitation density (or the filling fraction of dimers) is not pinned to exactly $1/4$ due to quantum fluctuations. Instead, the number of Rydberg excitations is now described by a \textit{distribution} $P(N_{\mathrm{exc}})$ (with mean $<$\,$N/4$) which acquires a width, as shown in Fig.~\ref{fig_4}(c) for a $20\times 4$ cylinder as an example. Using this information, we first construct the particle-number distribution for a system with $N=2160$ sites---that is nine times as large---using the procedure described in the Methods. Then, we generate classical dimer-monomer coverings according to this new distribution subject to the same (cylindrical) boundary conditions as the quantum system. This procedure captures the essence of the quantum fluctuations in a larger system. We emphasize that the reconstructed particle-number distribution is not the exact one of a 2160-site system but rather an approximation thereto; however, this distinction is irrelevant to the question we are asking, which is whether hyperuniformity survives in the presence of quantum fluctuations.

\begin{figure}[tb]
\includegraphics[width=\linewidth]{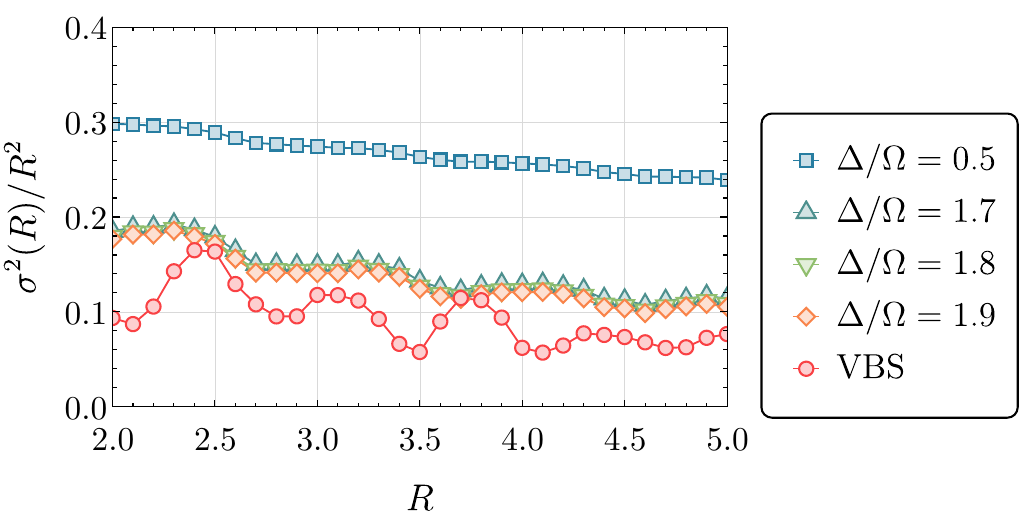} 
\caption{\label{fig_5}The scaled local number variance $\sigma^2(R)/R^2$ for the 2160-site classical analogs of the quantum systems at $\Delta/\Omega$ = 0.5, 1.7, 1.8 and 1.9 (the results are averaged over 10,000 configurations/snapshots) as well as for the ordered valence bond solid state.} 
\end{figure}

Our results in this regard are presented in Fig.~\ref{fig_5}, which shows the scaled local number variance $\sigma^2(R)/R^2$. Interestingly, for $\Delta/\Omega$ = 1.7, 1.8 and 1.9, and the VBS state at larger $\Delta/\Omega$, the local number variance $\sigma^2(R)$ increases more slowly than $R^2$ at large $R$, suggesting possible (effective) hyperuniformity of the corresponding systems. Note that for general 2D systems that may be class-I hyperuniform, the variance $\sigma^2(R)$ rigorously obeys the large-$R$ asymptotic relation $\sigma^2(R) = AR^2 +BR +o(R)$, where $A$ and $B$ are constants involving certain moments of the total correlation function, and $o(R)$ represents all terms of order less than $R$ \cite{To03}. For perfectly hyperuniform systems, $A=0$ and the ratio $B/A$ diverges; in general, as a system tends towards hyperuniformity, the ratio $B/A$ increases \cite{To21b}. Therefore, the ratio $B/A$ provides a measure of the degree of hyperuniformity \cite{To21b}, and here, we classify any system with $B/A \geq 10$ as effectively hyperuniform. In practice, to obtain smooth asymptotic behaviors, it is often advantageous to perform a quadratic fitting of the cumulative moving average $\overline{\sigma^2}(R)$ defined as $\overline{\sigma^2}(R) \equiv \int_0^R dx\, \sigma^2(x)/R$ instead of the variance $\sigma^2(R)$ \cite{To03, Ki17}. Using $\overline{\sigma^2}(R)$ within the range $2.0 \leq R \leq 5.0$, we compute the $B/A$ ratio to be 2.54, 10.10, 11.27, and 12.48 for $\Delta/\Omega$ = 0.5, 1.7, 1.8 and 1.9, respectively, suggesting that the three states in the QSL phase (at $\Delta/\Omega$ = 1.7, 1.8 and 1.9) are effectively hyperuniform, while the trivial state at $\Delta/\Omega = 0.5$ is far from being hyperuniform. The associated mean filling fractions are found to be 0.124, 0.231, 0.234, and 0.236 for $\Delta/\Omega$ = 0.5, 1.7, 1.8 and 1.9, respectively. On the other hand, the long-range-ordered VBS state, which is a pinwheel crystal \cite{nikolic2003physics,singh2007ground,poilblanc2011competing}, is not only perfectly hyperuniform, but also stealthy hyperuniform due to its periodic structure, i.e., its $S(k)$ is zero for an extended \textit{range} of small $k$ near the origin \cite{Ba08, To15}, in distinction to the QSL. These results clearly show that metrics based on the framework of hyperuniformity can be used to distinguish between the three distinct quantum phases of the Rydberg atom array. 

In addition, we find that the classical dimer-monomer coverings at a fixed filling fraction $f$ become effectively hyperuniform only as $f$ increases above 0.227 (see SI for details). For the QSL, however, $f$ is no longer fixed for different snapshots due to quantum fluctuations and the departure from perfect hyperuniformity is highly nonlinear as $f$ deviates from its maximum possible value. Hence, the hyperuniformity of quantum systems is further degraded compared to their classical analogs at the same (mean) filling fraction. For example, we find the quantum liquid to be effectively hyperuniform (i.e., $B/A\geq 10$) only for a mean filling fraction $f$ above $\simeq$ 0.231, which is 0.004 \textit{more} than the requirement for the classical dimer-monomer coverings.

\section{Discussion and outlook}
In this work, we uncover the hidden hyperuniformity of paradigmatic classical and quantum spin liquids, and demonstrate how such hyperuniformity can be used as a powerful tool to distinguish spin liquids from both trivial disordered (such as a quantum paramagnet) and crystalline ordered (e.g., valence bond solid) states. Specifically, we establish the perfect disordered hyperuniformity of classical ensembles of dimer coverings on different lattices as well as for the corresponding ideal quantum RVB states, which are $\mathbb{Z}_2$ QSLs. It is noteworthy that we find that $S(k)$ is analytic at the origin with small-$k$ scaling exponent $\alpha=6$. Interestingly, it has recently been shown that  $\alpha$ is also equal to 6 for the  quasiperiodic Penrose tiling, but there $S(k)$ is nonanalytic at the origin, implying a power-law decay in the pair correlation function \cite{Hi24}, to be contrasted with a compactly supported pair correlation function for the dimer covering. Moreover, we show that the $\mathbb{Z}_2$ QSL remains effectively hyperuniform even in the presence of a nonzero density of spinons and visons, which are inevitable in any realistic experimental setting. In contrast to the $\mathbb{Z}_2$ QSL, the trivial disordered state is far from being effectively hyperuniform, while the VBS state is perfectly stealthy hyperuniform.

We emphasize that hyperuniformity is a structural metric, so it cannot probe the signature ground state degeneracy on a torus that characterizes a topological phase. Moreover, hyperuniformity---being a measure of density fluctuations---is insensitive to phase perturbations and accordingly, to the existence in the spectrum of fractionalized vison excitations that distinguish a quantum spin liquid from a classical one. Nonetheless, hyperuniformity does tell us about the particular nature of the ``liquidity'' of CSLs and QSLs by quantifying the large-length-scale density fluctuations. The disordered hyperuniformity of CSLs and QSLs originates from the satisfaction of an underlying dimer constraint that results in fluctuations concentrated near the surface of an observation window, with the local number variance scaling linearly with $R$ at large $R$ (see the intuitive argument in Sec. 1 for details), in sharp distinction to a simple paramagnet. Hence, even though the trivial disordered and QSL phases both lack any local spatial order, they can be distinguished based on the unique fingerprints of each phase in the framework of hyperuniformity. Thus, our results suggest that one can first perform single-site measurements in experiments and then employ structural analysis that leverages metrics based on the framework of hyperuniformity to identify potential QSL candidates from these measurements, which is often a challenging task. After this initial identification, more advanced and computationally demanding quantum numerics can be employed to confirm the existence of a true QSL phase.


Our finding presents the $\mathbb{Z}_2$ quantum spin liquid as an example of hyperuniformity in a strongly interacting, highly entangled quantum system. The fact that we observe hyperuniformity in multiple lattice realizations of $\mathbb{Z}_2$ spin liquids suggests that hyperuniformity may be a generic feature of such QSLs, likely originating from the fundamentally constrained character of close-packed  dimer configurations (which, in turn, is intimately related to the emergent large-scale entanglement structure of these systems). However, the small-$k$ scaling exponent $\alpha$ differs for dimers on the kagome and triangular lattices, which is the result of the different microscopic dimer constraints. For example, in the kagome lattice, \textit{all} the first-, second-, and third-nearest-neighboring pairs of dimers are forbidden by the dimer constraint, whereas, for the triangular lattice, certain third-neighbor pairs of dimers are permitted. The dependence of $\alpha$ on the lattice structure as well as the theoretical determination of its exact value in the infinite-system limit for perfect dimer coverings on other lattices (generalizing the analysis presented in SI) will be the subject of our future investigation. 

It would also be interesting to extend our work to other QSLs (such as, e.g., RVB states constructed from trimer---rather than dimer---coverings \cite{PhysRevB.106.195155,kornjavca2022trimer}), which may have different invariant gauge groups but are still fundamentally characterized by long-range quantum entanglement. We expect to see hyperuniformity in other QSLs where some form of local regularity is present that leads to fluctuations concentrated near the surface of an observation window for a large window. For example, in a spin ice \cite{Sn01}, according to the ``ice rule'' or the ``2-in, 2-out rule'', in each tetrahedron,  two spins must point inward and two spins must point outward, and we expect the spatial distribution of ``in'' (or ``out'') spins to be hyperuniform. However, the implications of the exact form of local regularity for the hyperuniformity property of the associated spin liquids remains an open question that merits further exploration. It is noteworthy that our prediction of hyperuniformity also applies to quantum materials, such as NaYbSe$_2$, which was proposed as a candidate for a $\mathbb{Z}_2$ QSL \cite{Sc24b}. Moreover, a proximate spin liquid state was detected in KYbSe$_2$ using entanglement witnesses such as one tangle, two tangle, and quantum Fisher information \cite{Sc24}. However, this proximate state has recently been shown to possess long-range 120$^\circ$ magnetic order \cite{Sc24b}, so we expect it to be stealthy hyperuniform, unlike the effective nonstealthy hyperuniformity of the QSLs studied in this work.

\section*{Methods}
\subsection*{Generation of classical dimer coverings}
\begin{figure}[htb]
    \centering
    \includegraphics[width=\linewidth, trim={3.5cm 7.25cm 2.5cm 7cm}, clip]{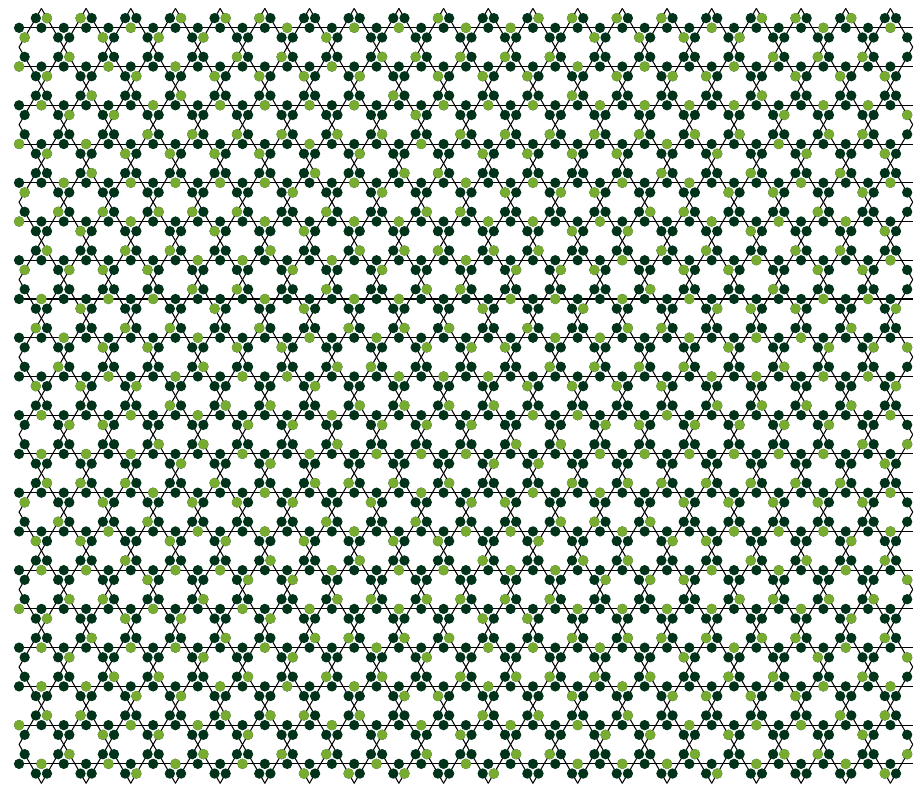}
    \caption{A configuration of $N=2400$ atoms satisfying the dimer constraint (see Fig. S5 in the SI for a visualization of just the dimers), as generated by our simulated annealing algorithm. As before, light (dark) green circles represent an atom in the $\vert r\rangle$ ($\vert g\rangle$) state.}
    \label{fig:fig_6}
\end{figure}

We devise a simulated annealing algorithm to obtain classical dimer coverings on the kagome and triangular lattices at various filling fractions $f$. Due to the singular nature of the hard-core exclusion potential between dimers, which makes it difficult to be directly incorporated into a conventional simulated annealing algorithm, we instead use the number of dimer pairs that violate the Rydberg blockade constraint as a fictitious energy $E$. For a given filling fraction $f$, we start by randomly assigning $fN$ bonds to be occupied by dimers and all the others unoccupied, where $N$ is the total number of bonds on the kagome lattice, or equivalently, the total number of lattices sites on the ruby lattice. Due to this random assignment, there are initially multiple violations of the hard-core exclusion between dimers. To resolve these violations (which is a nontrivial task, especially for $f \simeq 1/4$ and large $N$), at each time step, we randomly select an occupied edge with at least one of its nearest neighbors not occupied, move the dimer to one of its unoccupied neighboring edge, and accept the trial state swap according to the probability
\begin{equation}
\label{eq_5} p^{}_{\mathrm{acc}}(\mathrm{old}\rightarrow \mathrm{new}) = \textnormal{min}\left\{1, \textnormal{exp}\left(-\frac{E_{\mathrm{new}}-E_{\mathrm{old}}}{T}\right)\right\}.
\end{equation}
Here, $T$ is the fictitious temperature of the system that is initially set  high and gradually decreased according to a cooling schedule \cite{Ye98a, Ye98b}, and $E_{new}$ and $E_{old}$ are the fictitious energy associated with the configuration after and before the trial move, respectively. Note that in the kagome lattice, each edge shares a common vertex with 6 other edges, while in the triangular lattice each edge shares a common vertex with 10 other edges. We stop the simulation when we obtain a configuration with $E=0$, i.e., no violations of the Rydberg blockade are found. A representative example of a configuration of $N=2400$ atoms generated using this algorithm is shown in Fig.~\ref{fig:fig_6}. We emphasize here that the hard-core exclusion constraint at  $f=1/4$ is fully satisfied in this configuration, which is nontrivial to achieve for this relatively large system size. Using this algorithm, we are able to efficiently generate classical dimer coverings on the kagome lattice that fully satisfy the hard-core exclusion constraint for filling fractions up to the maximum $f=1/4$ and for system sizes as large as $N = 9600$.

\subsection*{Quantum numerics}

We study the Hamiltonian \eqref{eq:HRyd}, describing an interacting system of Rydberg atoms on the ruby lattice, using DMRG, which is a tensor-network algorithm that provides an optimized matrix product state representation of a target wavefunction. The DMRG calculations were preformed using the ITensor library \cite{ITensor}. Throughout our computations, we maintain a truncation error of $< 10^{-7}$ by adaptively increasing the bond dimension as required, up to bond dimension $\chi = 1600$. Employing cylindrical (periodic along the vertical direction and open along the horizontal) boundary conditions, we explore the possible ground states for a broad range of $\Delta/\Omega$. For small $\Delta/\Omega = 0.5$, we find that the ground state lies in the trivial disordered phase while it is a $\mathbb{Z}_2$ QSL for $\Delta/\Omega = 1.7,1.8,1.9$. 

To estimate the distribution of Rydberg excitations for a large 2160-site quantum system, we divide the lattice into nine 240-site subsystems in a 3-by-3 partition; the total number of Rydberg excitations $N_{\mathrm{exc},\mathrm{tot}}$ is given by
\begin{equation}
    N_{\mathrm{exc},\mathrm{tot}} = \sum_{i=1}^{6} N_{\mathrm{exc},1i} + \sum_{j=1}^{3} N_{\mathrm{exc},2j},
\end{equation}
where $N_{\mathrm{exc},1i}$ is the number of Rydberg excitations in the six subsystems in the left and right columns (i.e., those with one open boundary in the horizontal direction), while $N_{\mathrm{exc},2j}$ is the number of Rydberg excitations in the three subsystems in the central column (i.e., in the bulk or interior). This distinction between the boundary and the bulk is necessary since the density of Rydberg excitations is higher along the edges. We further assume that $N_{\mathrm{exc},1i}$ ($i=1,\ldots,6$) are independent and identically distributed (i.i.d.) random variables following the same distribution $P_1(N_{\mathrm{exc},1})$, which we estimate from 10000 snapshots of the leftmost or rightmost 240-site subregion of a 288-site quantum system simulated using DMRG---these subsystems therefore have one edge with open boundary conditions imposed. Similarly, we assume that $N_{\mathrm{exc},2j}$ ($j=1,2,3$) are i.i.d. random variables following the distribution $P_2(N_{\mathrm{exc},2})$ that we estimate from 10000 snapshots of a 240-site bulk region in the interior of a 288-site quantum system, which eliminates edge effects. Stitching these nine subsystems together provides a reasonable approximation to the number distribution of a 60$\times$12 system with 2160 lattice sites and cylindrical boundary conditions.

\begin{acknowledgments}
R.S. is supported by the Princeton Quantum Initiative Fellowship. Y. J. and S. T. are supported by the Army Research Office under Cooperative Agreement No. W911NF-22-2-0103. This work was performed in part at the Aspen Center for Physics, which is supported by National Science Foundation grant PHY-2210452. The participation of R.S. at the Aspen Center for Physics was supported by the Simons Foundation. The density-matrix renormalization group (DMRG) calculations presented in this paper were performed on computational resources managed and supported by Princeton Research Computing, a consortium of groups including the Princeton Institute for Computational Science and Engineering (PICSciE) and the Office of Information Technology's High Performance Computing Center and Visualization Laboratory at Princeton University.
\end{acknowledgments}


\begin{thebibliography}{115}%
\makeatletter
\providecommand \@ifxundefined [1]{%
 \@ifx{#1\undefined}
}%
\providecommand \@ifnum [1]{%
 \ifnum #1\expandafter \@firstoftwo
 \else \expandafter \@secondoftwo
 \fi
}%
\providecommand \@ifx [1]{%
 \ifx #1\expandafter \@firstoftwo
 \else \expandafter \@secondoftwo
 \fi
}%
\providecommand \natexlab [1]{#1}%
\providecommand \enquote  [1]{``#1''}%
\providecommand \bibnamefont  [1]{#1}%
\providecommand \bibfnamefont [1]{#1}%
\providecommand \citenamefont [1]{#1}%
\providecommand \href@noop [0]{\@secondoftwo}%
\providecommand \href [0]{\begingroup \@sanitize@url \@href}%
\providecommand \@href[1]{\@@startlink{#1}\@@href}%
\providecommand \@@href[1]{\endgroup#1\@@endlink}%
\providecommand \@sanitize@url [0]{\catcode `\\12\catcode `\$12\catcode `\&12\catcode `\#12\catcode `\^12\catcode `\_12\catcode `\%12\relax}%
\providecommand \@@startlink[1]{}%
\providecommand \@@endlink[0]{}%
\providecommand \url  [0]{\begingroup\@sanitize@url \@url }%
\providecommand \@url [1]{\endgroup\@href {#1}{\urlprefix }}%
\providecommand \urlprefix  [0]{URL }%
\providecommand \Eprint [0]{\href }%
\providecommand \doibase [0]{https://doi.org/}%
\providecommand \selectlanguage [0]{\@gobble}%
\providecommand \bibinfo  [0]{\@secondoftwo}%
\providecommand \bibfield  [0]{\@secondoftwo}%
\providecommand \translation [1]{[#1]}%
\providecommand \BibitemOpen [0]{}%
\providecommand \bibitemStop [0]{}%
\providecommand \bibitemNoStop [0]{.\EOS\space}%
\providecommand \EOS [0]{\spacefactor3000\relax}%
\providecommand \BibitemShut  [1]{\csname bibitem#1\endcsname}%
\let\auto@bib@innerbib\@empty
\bibitem [{\citenamefont {Torquato}\ and\ \citenamefont {Stillinger}(2003)}]{To03}%
  \BibitemOpen
  \bibfield  {author} {\bibinfo {author} {\bibfnamefont {S.}~\bibnamefont {Torquato}}\ and\ \bibinfo {author} {\bibfnamefont {F.~H.}\ \bibnamefont {Stillinger}},\ }\bibfield  {title} {\bibinfo {title} {Local density fluctuations, hyperuniformity, and order metrics},\ }\href {https://doi.org/https://doi.org/10.1103/PhysRevE.68.041113} {\bibfield  {journal} {\bibinfo  {journal} {Phys. Rev. E}\ }\textbf {\bibinfo {volume} {68}},\ \bibinfo {pages} {041113} (\bibinfo {year} {2003})}\BibitemShut {NoStop}%
\bibitem [{\citenamefont {Torquato}(2018)}]{To18a}%
  \BibitemOpen
  \bibfield  {author} {\bibinfo {author} {\bibfnamefont {S.}~\bibnamefont {Torquato}},\ }\bibfield  {title} {\bibinfo {title} {Hyperuniform states of matter},\ }\href {https://doi.org/https://doi.org/10.1016/j.physrep.2018.03.001} {\bibfield  {journal} {\bibinfo  {journal} {Phys. Rep.}\ }\textbf {\bibinfo {volume} {745}},\ \bibinfo {pages} {1} (\bibinfo {year} {2018})}\BibitemShut {NoStop}%
\bibitem [{\citenamefont {Zachary}\ and\ \citenamefont {Torquato}(2009)}]{Za09}%
  \BibitemOpen
  \bibfield  {author} {\bibinfo {author} {\bibfnamefont {C.~E.}\ \bibnamefont {Zachary}}\ and\ \bibinfo {author} {\bibfnamefont {S.}~\bibnamefont {Torquato}},\ }\bibfield  {title} {\bibinfo {title} {Hyperuniformity in point patterns and two-phase random heterogeneous media},\ }\href {https://doi.org/https://doi.org/10.1088/1742-5468/2009/12/P12015} {\bibfield  {journal} {\bibinfo  {journal} {J. Stat. Mech. Theor. Exp.}\ }\textbf {\bibinfo {volume} {2009}},\ \bibinfo {pages} {P12015} (\bibinfo {year} {2009})}\BibitemShut {NoStop}%
\bibitem [{\citenamefont {Jiao}\ \emph {et~al.}(2014)\citenamefont {Jiao}, \citenamefont {Lau}, \citenamefont {Hatzikirou}, \citenamefont {Meyer-Hermann}, \citenamefont {Corbo},\ and\ \citenamefont {Torquato}}]{Ji14}%
  \BibitemOpen
  \bibfield  {author} {\bibinfo {author} {\bibfnamefont {Y.}~\bibnamefont {Jiao}}, \bibinfo {author} {\bibfnamefont {T.}~\bibnamefont {Lau}}, \bibinfo {author} {\bibfnamefont {H.}~\bibnamefont {Hatzikirou}}, \bibinfo {author} {\bibfnamefont {M.}~\bibnamefont {Meyer-Hermann}}, \bibinfo {author} {\bibfnamefont {J.~C.}\ \bibnamefont {Corbo}},\ and\ \bibinfo {author} {\bibfnamefont {S.}~\bibnamefont {Torquato}},\ }\bibfield  {title} {\bibinfo {title} {Avian photoreceptor patterns represent a disordered hyperuniform solution to a multiscale packing problem},\ }\href {https://doi.org/https://doi.org/10.1103/PhysRevE.89.022721} {\bibfield  {journal} {\bibinfo  {journal} {Phys. Rev. E}\ }\textbf {\bibinfo {volume} {89}},\ \bibinfo {pages} {022721} (\bibinfo {year} {2014})}\BibitemShut {NoStop}%
\bibitem [{\citenamefont {Dreyfus}\ \emph {et~al.}(2015)\citenamefont {Dreyfus}, \citenamefont {Xu}, \citenamefont {Still}, \citenamefont {Hough}, \citenamefont {Yodh},\ and\ \citenamefont {Torquato}}]{Dr15}%
  \BibitemOpen
  \bibfield  {author} {\bibinfo {author} {\bibfnamefont {R.}~\bibnamefont {Dreyfus}}, \bibinfo {author} {\bibfnamefont {Y.}~\bibnamefont {Xu}}, \bibinfo {author} {\bibfnamefont {T.}~\bibnamefont {Still}}, \bibinfo {author} {\bibfnamefont {L.~A.}\ \bibnamefont {Hough}}, \bibinfo {author} {\bibfnamefont {A.~G.}\ \bibnamefont {Yodh}},\ and\ \bibinfo {author} {\bibfnamefont {S.}~\bibnamefont {Torquato}},\ }\bibfield  {title} {\bibinfo {title} {Diagnosing hyperuniformity in two-dimensional, disordered, jammed packings of soft spheres},\ }\href {https://doi.org/https://doi.org/10.1103/PhysRevE.91.012302} {\bibfield  {journal} {\bibinfo  {journal} {Phys. Rev. E}\ }\textbf {\bibinfo {volume} {91}},\ \bibinfo {pages} {012302} (\bibinfo {year} {2015})}\BibitemShut {NoStop}%
\bibitem [{\citenamefont {Hexner}\ and\ \citenamefont {Levine}(2015)}]{He15}%
  \BibitemOpen
  \bibfield  {author} {\bibinfo {author} {\bibfnamefont {D.}~\bibnamefont {Hexner}}\ and\ \bibinfo {author} {\bibfnamefont {D.}~\bibnamefont {Levine}},\ }\bibfield  {title} {\bibinfo {title} {Hyperuniformity of critical absorbing states},\ }\href {https://doi.org/https://doi.org/10.1103/PhysRevLett.114.110602} {\bibfield  {journal} {\bibinfo  {journal} {Phys. Rev. Lett.}\ }\textbf {\bibinfo {volume} {114}},\ \bibinfo {pages} {110602} (\bibinfo {year} {2015})}\BibitemShut {NoStop}%
\bibitem [{\citenamefont {Jack}\ \emph {et~al.}(2015)\citenamefont {Jack}, \citenamefont {Thompson},\ and\ \citenamefont {Sollich}}]{Ja15}%
  \BibitemOpen
  \bibfield  {author} {\bibinfo {author} {\bibfnamefont {R.~L.}\ \bibnamefont {Jack}}, \bibinfo {author} {\bibfnamefont {I.~R.}\ \bibnamefont {Thompson}},\ and\ \bibinfo {author} {\bibfnamefont {P.}~\bibnamefont {Sollich}},\ }\bibfield  {title} {\bibinfo {title} {Hyperuniformity and phase separation in biased ensembles of trajectories for diffusive systems},\ }\href {https://doi.org/https://doi.org/10.1103/PhysRevLett.114.060601} {\bibfield  {journal} {\bibinfo  {journal} {Phys. Rev. Lett.}\ }\textbf {\bibinfo {volume} {114}},\ \bibinfo {pages} {060601} (\bibinfo {year} {2015})}\BibitemShut {NoStop}%
\bibitem [{\citenamefont {Rumi}\ \emph {et~al.}(2019)\citenamefont {Rumi}, \citenamefont {S{\'a}nchez}, \citenamefont {El{\'\i}as}, \citenamefont {Maldonado}, \citenamefont {Puig}, \citenamefont {Bolecek}, \citenamefont {Nieva}, \citenamefont {Konczykowski}, \citenamefont {Fasano},\ and\ \citenamefont {Kolton}}]{Ru19}%
  \BibitemOpen
  \bibfield  {author} {\bibinfo {author} {\bibfnamefont {G.}~\bibnamefont {Rumi}}, \bibinfo {author} {\bibfnamefont {J.~A.}\ \bibnamefont {S{\'a}nchez}}, \bibinfo {author} {\bibfnamefont {F.}~\bibnamefont {El{\'\i}as}}, \bibinfo {author} {\bibfnamefont {R.~C.}\ \bibnamefont {Maldonado}}, \bibinfo {author} {\bibfnamefont {J.}~\bibnamefont {Puig}}, \bibinfo {author} {\bibfnamefont {N.~R.~C.}\ \bibnamefont {Bolecek}}, \bibinfo {author} {\bibfnamefont {G.}~\bibnamefont {Nieva}}, \bibinfo {author} {\bibfnamefont {M.}~\bibnamefont {Konczykowski}}, \bibinfo {author} {\bibfnamefont {Y.}~\bibnamefont {Fasano}},\ and\ \bibinfo {author} {\bibfnamefont {A.~B.}\ \bibnamefont {Kolton}},\ }\bibfield  {title} {\bibinfo {title} {Hyperuniform vortex patterns at the surface of type-ii superconductors},\ }\href {https://doi.org/https://doi.org/10.1103/PhysRevResearch.1.033057} {\bibfield  {journal} {\bibinfo  {journal} {Phys. Rev. Res.}\ }\textbf {\bibinfo {volume} {1}},\ \bibinfo {pages} {033057} (\bibinfo {year}
  {2019})}\BibitemShut {NoStop}%
\bibitem [{\citenamefont {Lei}\ \emph {et~al.}(2019)\citenamefont {Lei}, \citenamefont {Ciamarra},\ and\ \citenamefont {Ni}}]{Le19a}%
  \BibitemOpen
  \bibfield  {author} {\bibinfo {author} {\bibfnamefont {Q.-L.}\ \bibnamefont {Lei}}, \bibinfo {author} {\bibfnamefont {M.~P.}\ \bibnamefont {Ciamarra}},\ and\ \bibinfo {author} {\bibfnamefont {R.}~\bibnamefont {Ni}},\ }\bibfield  {title} {\bibinfo {title} {Nonequilibrium strongly hyperuniform fluids of circle active particles with large local density fluctuations},\ }\href {https://doi.org/https://doi.org/10.1126/sciadv.aau7423} {\bibfield  {journal} {\bibinfo  {journal} {Sci. Adv.}\ }\textbf {\bibinfo {volume} {5}},\ \bibinfo {pages} {eaau7423} (\bibinfo {year} {2019})}\BibitemShut {NoStop}%
\bibitem [{\citenamefont {Lei}\ and\ \citenamefont {Ni}(2019)}]{Le19b}%
  \BibitemOpen
  \bibfield  {author} {\bibinfo {author} {\bibfnamefont {Q.-L.}\ \bibnamefont {Lei}}\ and\ \bibinfo {author} {\bibfnamefont {R.}~\bibnamefont {Ni}},\ }\bibfield  {title} {\bibinfo {title} {Hydrodynamics of random-organizing hyperuniform fluids},\ }\href {https://doi.org/https://doi.org/10.1073/pnas.1911596116} {\bibfield  {journal} {\bibinfo  {journal} {Proc. Natl. Acad. Sci. U.S.A.}\ }\textbf {\bibinfo {volume} {116}},\ \bibinfo {pages} {22983} (\bibinfo {year} {2019})}\BibitemShut {NoStop}%
\bibitem [{\citenamefont {Huang}\ \emph {et~al.}(2021)\citenamefont {Huang}, \citenamefont {Hu}, \citenamefont {Yang}, \citenamefont {Liu},\ and\ \citenamefont {Zhang}}]{Hu21}%
  \BibitemOpen
  \bibfield  {author} {\bibinfo {author} {\bibfnamefont {M.}~\bibnamefont {Huang}}, \bibinfo {author} {\bibfnamefont {W.}~\bibnamefont {Hu}}, \bibinfo {author} {\bibfnamefont {S.}~\bibnamefont {Yang}}, \bibinfo {author} {\bibfnamefont {Q.-X.}\ \bibnamefont {Liu}},\ and\ \bibinfo {author} {\bibfnamefont {H.~P.}\ \bibnamefont {Zhang}},\ }\bibfield  {title} {\bibinfo {title} {Circular swimming motility and disordered hyperuniform state in an algae system},\ }\href {https://doi.org/https://doi.org/10.1073/pnas.2100493118} {\bibfield  {journal} {\bibinfo  {journal} {Proc. Natl. Acad. Sci. U.S.A.}\ }\textbf {\bibinfo {volume} {118}},\ \bibinfo {pages} {e2100493118} (\bibinfo {year} {2021})}\BibitemShut {NoStop}%
\bibitem [{\citenamefont {Liu}\ \emph {et~al.}(2024)\citenamefont {Liu}, \citenamefont {Chen}, \citenamefont {Tian}, \citenamefont {Xu},\ and\ \citenamefont {Jiao}}]{Li24}%
  \BibitemOpen
  \bibfield  {author} {\bibinfo {author} {\bibfnamefont {Y.}~\bibnamefont {Liu}}, \bibinfo {author} {\bibfnamefont {D.}~\bibnamefont {Chen}}, \bibinfo {author} {\bibfnamefont {J.}~\bibnamefont {Tian}}, \bibinfo {author} {\bibfnamefont {W.}~\bibnamefont {Xu}},\ and\ \bibinfo {author} {\bibfnamefont {Y.}~\bibnamefont {Jiao}},\ }\bibfield  {title} {\bibinfo {title} {Universal hyperuniform organization in looped leaf vein networks},\ }\href {https://doi.org/10.1103/PhysRevLett.133.028401} {\bibfield  {journal} {\bibinfo  {journal} {Phys. Rev. Lett.}\ }\textbf {\bibinfo {volume} {133}},\ \bibinfo {pages} {028401} (\bibinfo {year} {2024})}\BibitemShut {NoStop}%
\bibitem [{\citenamefont {Florescu}\ \emph {et~al.}(2009)\citenamefont {Florescu}, \citenamefont {Torquato},\ and\ \citenamefont {Steinhardt}}]{Fl09}%
  \BibitemOpen
  \bibfield  {author} {\bibinfo {author} {\bibfnamefont {M.}~\bibnamefont {Florescu}}, \bibinfo {author} {\bibfnamefont {S.}~\bibnamefont {Torquato}},\ and\ \bibinfo {author} {\bibfnamefont {P.~J.}\ \bibnamefont {Steinhardt}},\ }\bibfield  {title} {\bibinfo {title} {Designer disordered materials with large, complete photonic band gaps},\ }\href {https://doi.org/https://doi.org/10.1073/pnas.0907744106} {\bibfield  {journal} {\bibinfo  {journal} {Proc. Natl. Acad. Sci. U.S.A.}\ }\textbf {\bibinfo {volume} {106}},\ \bibinfo {pages} {20658} (\bibinfo {year} {2009})}\BibitemShut {NoStop}%
\bibitem [{\citenamefont {Man}\ \emph {et~al.}(2013)\citenamefont {Man}, \citenamefont {Florescu}, \citenamefont {Williamson}, \citenamefont {He}, \citenamefont {Hashemizad}, \citenamefont {Leung}, \citenamefont {Liner}, \citenamefont {Torquato}, \citenamefont {Chaikin},\ and\ \citenamefont {Steinhardt}}]{Ma13}%
  \BibitemOpen
  \bibfield  {author} {\bibinfo {author} {\bibfnamefont {W.}~\bibnamefont {Man}}, \bibinfo {author} {\bibfnamefont {M.}~\bibnamefont {Florescu}}, \bibinfo {author} {\bibfnamefont {E.~P.}\ \bibnamefont {Williamson}}, \bibinfo {author} {\bibfnamefont {Y.}~\bibnamefont {He}}, \bibinfo {author} {\bibfnamefont {S.~R.}\ \bibnamefont {Hashemizad}}, \bibinfo {author} {\bibfnamefont {B.~Y.~C.}\ \bibnamefont {Leung}}, \bibinfo {author} {\bibfnamefont {D.~R.}\ \bibnamefont {Liner}}, \bibinfo {author} {\bibfnamefont {S.}~\bibnamefont {Torquato}}, \bibinfo {author} {\bibfnamefont {P.~M.}\ \bibnamefont {Chaikin}},\ and\ \bibinfo {author} {\bibfnamefont {P.~J.}\ \bibnamefont {Steinhardt}},\ }\bibfield  {title} {\bibinfo {title} {Isotropic band gaps and freeform waveguides observed in hyperuniform disordered photonic solids},\ }\href {https://doi.org/https://doi.org/10.1073/pnas.130787911} {\bibfield  {journal} {\bibinfo  {journal} {Proc. Natl. Acad. Sci. U.S.A.}\ }\textbf {\bibinfo {volume} {110}},\ \bibinfo {pages} {15886}
  (\bibinfo {year} {2013})}\BibitemShut {NoStop}%
\bibitem [{\citenamefont {Leseur}\ \emph {et~al.}(2016)\citenamefont {Leseur}, \citenamefont {Pierrat},\ and\ \citenamefont {Carminati}}]{Le16}%
  \BibitemOpen
  \bibfield  {author} {\bibinfo {author} {\bibfnamefont {O.}~\bibnamefont {Leseur}}, \bibinfo {author} {\bibfnamefont {R.}~\bibnamefont {Pierrat}},\ and\ \bibinfo {author} {\bibfnamefont {R.}~\bibnamefont {Carminati}},\ }\bibfield  {title} {\bibinfo {title} {High-density hyperuniform materials can be transparent},\ }\href {https://doi.org/https://doi.org/10.1364/OPTICA.3.000763} {\bibfield  {journal} {\bibinfo  {journal} {Optica}\ }\textbf {\bibinfo {volume} {3}},\ \bibinfo {pages} {763} (\bibinfo {year} {2016})}\BibitemShut {NoStop}%
\bibitem [{\citenamefont {Zhang}\ \emph {et~al.}(2016)\citenamefont {Zhang}, \citenamefont {Stillinger},\ and\ \citenamefont {Torquato}}]{Zh16}%
  \BibitemOpen
  \bibfield  {author} {\bibinfo {author} {\bibfnamefont {G.}~\bibnamefont {Zhang}}, \bibinfo {author} {\bibfnamefont {F.~H.}\ \bibnamefont {Stillinger}},\ and\ \bibinfo {author} {\bibfnamefont {S.}~\bibnamefont {Torquato}},\ }\bibfield  {title} {\bibinfo {title} {Transport, geometrical, and topological properties of stealthy disordered hyperuniform two-phase systems},\ }\href {https://doi.org/https://doi.org/10.1063/1.4972862} {\bibfield  {journal} {\bibinfo  {journal} {J. Chem. Phys.}\ }\textbf {\bibinfo {volume} {145}},\ \bibinfo {pages} {244109} (\bibinfo {year} {2016})}\BibitemShut {NoStop}%
\bibitem [{\citenamefont {Chen}\ and\ \citenamefont {Torquato}(2018)}]{Ch18a}%
  \BibitemOpen
  \bibfield  {author} {\bibinfo {author} {\bibfnamefont {D.}~\bibnamefont {Chen}}\ and\ \bibinfo {author} {\bibfnamefont {S.}~\bibnamefont {Torquato}},\ }\bibfield  {title} {\bibinfo {title} {Designing disordered hyperuniform two-phase materials with novel physical properties},\ }\href {https://doi.org/https://doi.org/10.1016/j.actamat.2017.09.053} {\bibfield  {journal} {\bibinfo  {journal} {Acta Mater.}\ }\textbf {\bibinfo {volume} {142}},\ \bibinfo {pages} {152} (\bibinfo {year} {2018})}\BibitemShut {NoStop}%
\bibitem [{\citenamefont {Xu}\ \emph {et~al.}(2017)\citenamefont {Xu}, \citenamefont {Chen}, \citenamefont {Chen}, \citenamefont {Xu},\ and\ \citenamefont {Jiao}}]{Xu17}%
  \BibitemOpen
  \bibfield  {author} {\bibinfo {author} {\bibfnamefont {Y.}~\bibnamefont {Xu}}, \bibinfo {author} {\bibfnamefont {S.}~\bibnamefont {Chen}}, \bibinfo {author} {\bibfnamefont {P.}~\bibnamefont {Chen}}, \bibinfo {author} {\bibfnamefont {W.}~\bibnamefont {Xu}},\ and\ \bibinfo {author} {\bibfnamefont {Y.}~\bibnamefont {Jiao}},\ }\bibfield  {title} {\bibinfo {title} {Microstructure and mechanical properties of hyperuniform heterogeneous materials},\ }\href {https://doi.org/https://doi.org/10.1103/PhysRevE.96.043301} {\bibfield  {journal} {\bibinfo  {journal} {Phys. Rev. E}\ }\textbf {\bibinfo {volume} {96}},\ \bibinfo {pages} {043301} (\bibinfo {year} {2017})}\BibitemShut {NoStop}%
\bibitem [{\citenamefont {Kim}\ and\ \citenamefont {Torquato}(2020)}]{Ki20}%
  \BibitemOpen
  \bibfield  {author} {\bibinfo {author} {\bibfnamefont {J.}~\bibnamefont {Kim}}\ and\ \bibinfo {author} {\bibfnamefont {S.}~\bibnamefont {Torquato}},\ }\bibfield  {title} {\bibinfo {title} {Multifunctional composites for elastic and electromagnetic wave propagation},\ }\href {https://doi.org/https://doi.org/10.1073/pnas.1914086117} {\bibfield  {journal} {\bibinfo  {journal} {Proc. Natl. Acad. Sci. U.S.A.}\ }\textbf {\bibinfo {volume} {117}},\ \bibinfo {pages} {8764} (\bibinfo {year} {2020})}\BibitemShut {NoStop}%
\bibitem [{\citenamefont {Kim}\ and\ \citenamefont {Torquato}(2023)}]{Ki23}%
  \BibitemOpen
  \bibfield  {author} {\bibinfo {author} {\bibfnamefont {J.}~\bibnamefont {Kim}}\ and\ \bibinfo {author} {\bibfnamefont {S.}~\bibnamefont {Torquato}},\ }\bibfield  {title} {\bibinfo {title} {Effective electromagnetic wave properties of disordered stealthy hyperuniform layered media beyond the quasistatic regime},\ }\href {https://doi.org/https://doi.org/10.1364/OPTICA.489797} {\bibfield  {journal} {\bibinfo  {journal} {Optica}\ }\textbf {\bibinfo {volume} {10}},\ \bibinfo {pages} {965} (\bibinfo {year} {2023})}\BibitemShut {NoStop}%
\bibitem [{\citenamefont {Klatt}\ \emph {et~al.}(2022)\citenamefont {Klatt}, \citenamefont {Steinhardt},\ and\ \citenamefont {Torquato}}]{Kl22}%
  \BibitemOpen
  \bibfield  {author} {\bibinfo {author} {\bibfnamefont {M.~A.}\ \bibnamefont {Klatt}}, \bibinfo {author} {\bibfnamefont {P.~J.}\ \bibnamefont {Steinhardt}},\ and\ \bibinfo {author} {\bibfnamefont {S.}~\bibnamefont {Torquato}},\ }\bibfield  {title} {\bibinfo {title} {Wave propagation and band tails of two-dimensional disordered systems in the thermodynamic limit},\ }\href {https://doi.org/https://doi.org/10.1073/pnas.2213633119} {\bibfield  {journal} {\bibinfo  {journal} {Proc. Natl. Acad. Sci. U.S.A.}\ }\textbf {\bibinfo {volume} {119}},\ \bibinfo {pages} {e2213633119} (\bibinfo {year} {2022})}\BibitemShut {NoStop}%
\bibitem [{\citenamefont {Aubry}\ \emph {et~al.}(2020)\citenamefont {Aubry}, \citenamefont {Froufe-P{\'e}rez}, \citenamefont {Kuhl}, \citenamefont {Legrand}, \citenamefont {Scheffold},\ and\ \citenamefont {Mortessagne}}]{Au20}%
  \BibitemOpen
  \bibfield  {author} {\bibinfo {author} {\bibfnamefont {G.~J.}\ \bibnamefont {Aubry}}, \bibinfo {author} {\bibfnamefont {L.~S.}\ \bibnamefont {Froufe-P{\'e}rez}}, \bibinfo {author} {\bibfnamefont {U.}~\bibnamefont {Kuhl}}, \bibinfo {author} {\bibfnamefont {O.}~\bibnamefont {Legrand}}, \bibinfo {author} {\bibfnamefont {F.}~\bibnamefont {Scheffold}},\ and\ \bibinfo {author} {\bibfnamefont {F.}~\bibnamefont {Mortessagne}},\ }\bibfield  {title} {\bibinfo {title} {Experimental tuning of transport regimes in hyperuniform disordered photonic materials},\ }\href {https://doi.org/https://doi.org/10.1103/PhysRevLett.125.127402} {\bibfield  {journal} {\bibinfo  {journal} {Phys. Rev. Lett.}\ }\textbf {\bibinfo {volume} {125}},\ \bibinfo {pages} {127402} (\bibinfo {year} {2020})}\BibitemShut {NoStop}%
\bibitem [{\citenamefont {Romero-Garc{\'\i}a}\ \emph {et~al.}(2021)\citenamefont {Romero-Garc{\'\i}a}, \citenamefont {Ch{\'e}ron}, \citenamefont {Kuznetsova}, \citenamefont {Groby}, \citenamefont {F{\'e}lix}, \citenamefont {Pagneux},\ and\ \citenamefont {Garcia-Raffi}}]{Ro21}%
  \BibitemOpen
  \bibfield  {author} {\bibinfo {author} {\bibfnamefont {V.}~\bibnamefont {Romero-Garc{\'\i}a}}, \bibinfo {author} {\bibfnamefont {{\'E}.}~\bibnamefont {Ch{\'e}ron}}, \bibinfo {author} {\bibfnamefont {S.}~\bibnamefont {Kuznetsova}}, \bibinfo {author} {\bibfnamefont {J.-P.}\ \bibnamefont {Groby}}, \bibinfo {author} {\bibfnamefont {S.}~\bibnamefont {F{\'e}lix}}, \bibinfo {author} {\bibfnamefont {V.}~\bibnamefont {Pagneux}},\ and\ \bibinfo {author} {\bibfnamefont {L.}~\bibnamefont {Garcia-Raffi}},\ }\bibfield  {title} {\bibinfo {title} {Wave transport in 1D stealthy hyperuniform phononic materials made of non-resonant and resonant scatterers},\ }\href {https://doi.org/https://doi.org/10.1063/5.0059928} {\bibfield  {journal} {\bibinfo  {journal} {APL Mater.}\ }\textbf {\bibinfo {volume} {9}},\ \bibinfo {pages} {101101} (\bibinfo {year} {2021})}\BibitemShut {NoStop}%
\bibitem [{\citenamefont {Aeby}\ \emph {et~al.}(2022)\citenamefont {Aeby}, \citenamefont {Aubry}, \citenamefont {Froufe-P{\'e}rez},\ and\ \citenamefont {Scheffold}}]{Ae22}%
  \BibitemOpen
  \bibfield  {author} {\bibinfo {author} {\bibfnamefont {S.}~\bibnamefont {Aeby}}, \bibinfo {author} {\bibfnamefont {G.~J.}\ \bibnamefont {Aubry}}, \bibinfo {author} {\bibfnamefont {L.~S.}\ \bibnamefont {Froufe-P{\'e}rez}},\ and\ \bibinfo {author} {\bibfnamefont {F.}~\bibnamefont {Scheffold}},\ }\bibfield  {title} {\bibinfo {title} {Fabrication of hyperuniform dielectric networks via heat-induced shrinkage reveals a bandgap at telecom wavelengths},\ }\href {https://doi.org/https://doi.org/10.1002/adom.202200232} {\bibfield  {journal} {\bibinfo  {journal} {Adv. Opt. Mater.}\ }\textbf {\bibinfo {volume} {10}},\ \bibinfo {pages} {2200232} (\bibinfo {year} {2022})}\BibitemShut {NoStop}%
\bibitem [{\citenamefont {Christogeorgos}\ \emph {et~al.}(2021)\citenamefont {Christogeorgos}, \citenamefont {Zhang}, \citenamefont {Cheng},\ and\ \citenamefont {Hao}}]{Ch22}%
  \BibitemOpen
  \bibfield  {author} {\bibinfo {author} {\bibfnamefont {O.}~\bibnamefont {Christogeorgos}}, \bibinfo {author} {\bibfnamefont {H.}~\bibnamefont {Zhang}}, \bibinfo {author} {\bibfnamefont {Q.}~\bibnamefont {Cheng}},\ and\ \bibinfo {author} {\bibfnamefont {Y.}~\bibnamefont {Hao}},\ }\bibfield  {title} {\bibinfo {title} {Extraordinary directive emission and scanning from an array of radiation sources with hyperuniform disorder},\ }\href {https://doi.org/https://doi.org/10.1103/PhysRevApplied.15.014062} {\bibfield  {journal} {\bibinfo  {journal} {Phys. Rev. Appl.}\ }\textbf {\bibinfo {volume} {15}},\ \bibinfo {pages} {014062} (\bibinfo {year} {2021})}\BibitemShut {NoStop}%
\bibitem [{\citenamefont {Tang}\ \emph {et~al.}(2022)\citenamefont {Tang}, \citenamefont {Hao}, \citenamefont {Liu}, \citenamefont {Tian}, \citenamefont {Niu},\ and\ \citenamefont {Zang}}]{Ta22}%
  \BibitemOpen
  \bibfield  {author} {\bibinfo {author} {\bibfnamefont {H.}~\bibnamefont {Tang}}, \bibinfo {author} {\bibfnamefont {Z.}~\bibnamefont {Hao}}, \bibinfo {author} {\bibfnamefont {Y.}~\bibnamefont {Liu}}, \bibinfo {author} {\bibfnamefont {Y.}~\bibnamefont {Tian}}, \bibinfo {author} {\bibfnamefont {H.}~\bibnamefont {Niu}},\ and\ \bibinfo {author} {\bibfnamefont {J.}~\bibnamefont {Zang}},\ }\bibfield  {title} {\bibinfo {title} {Soft and disordered hyperuniform elastic metamaterials for highly efficient vibration concentration},\ }\href {https://doi.org/https://doi.org/10.1093/nsr/nwab133} {\bibfield  {journal} {\bibinfo  {journal} {Natl. Sci. Rev.}\ }\textbf {\bibinfo {volume} {9}},\ \bibinfo {pages} {nwab133} (\bibinfo {year} {2022})}\BibitemShut {NoStop}%
\bibitem [{\citenamefont {Torquato}\ \emph {et~al.}(2008)\citenamefont {Torquato}, \citenamefont {Scardicchio},\ and\ \citenamefont {Zachary}}]{To08}%
  \BibitemOpen
  \bibfield  {author} {\bibinfo {author} {\bibfnamefont {S.}~\bibnamefont {Torquato}}, \bibinfo {author} {\bibfnamefont {A.}~\bibnamefont {Scardicchio}},\ and\ \bibinfo {author} {\bibfnamefont {C.~E.}\ \bibnamefont {Zachary}},\ }\bibfield  {title} {\bibinfo {title} {Point processes in arbitrary dimension from fermionic gases, random matrix theory, and number theory},\ }\href {https://doi.org/10.1088/1742-5468/2008/11/P11019} {\bibfield  {journal} {\bibinfo  {journal} {J. Stat. Mech. Theor. Exp.}\ }\textbf {\bibinfo {volume} {2008}},\ \bibinfo {pages} {P11019} (\bibinfo {year} {2008})}\BibitemShut {NoStop}%
\bibitem [{\citenamefont {Scardicchio}\ \emph {et~al.}(2009)\citenamefont {Scardicchio}, \citenamefont {Zachary},\ and\ \citenamefont {Torquato}}]{Sc09}%
  \BibitemOpen
  \bibfield  {author} {\bibinfo {author} {\bibfnamefont {A.}~\bibnamefont {Scardicchio}}, \bibinfo {author} {\bibfnamefont {C.~E.}\ \bibnamefont {Zachary}},\ and\ \bibinfo {author} {\bibfnamefont {S.}~\bibnamefont {Torquato}},\ }\bibfield  {title} {\bibinfo {title} {{Statistical properties of determinantal point processes in high-dimensional Euclidean spaces}},\ }\href {https://doi.org/10.1103/PhysRevE.79.041108} {\bibfield  {journal} {\bibinfo  {journal} {Phys. Rev. E}\ }\textbf {\bibinfo {volume} {79}},\ \bibinfo {pages} {041108} (\bibinfo {year} {2009})}\BibitemShut {NoStop}%
\bibitem [{\citenamefont {Abreu}\ \emph {et~al.}(2017)\citenamefont {Abreu}, \citenamefont {Pereira}, \citenamefont {Romero},\ and\ \citenamefont {Torquato}}]{Ab17}%
  \BibitemOpen
  \bibfield  {author} {\bibinfo {author} {\bibfnamefont {L.~D.}\ \bibnamefont {Abreu}}, \bibinfo {author} {\bibfnamefont {J.~M.}\ \bibnamefont {Pereira}}, \bibinfo {author} {\bibfnamefont {J.~L.}\ \bibnamefont {Romero}},\ and\ \bibinfo {author} {\bibfnamefont {S.}~\bibnamefont {Torquato}},\ }\bibfield  {title} {\bibinfo {title} {{The Weyl--Heisenberg ensemble: hyperuniformity and higher Landau levels}},\ }\href {https://doi.org/10.1088/1742-5468/aa68a7} {\bibfield  {journal} {\bibinfo  {journal} {J. Stat. Mech.: Theory Exp.}\ }\textbf {\bibinfo {volume} {2017}}\bibinfo  {number} { (4)},\ \bibinfo {pages} {043103}}\BibitemShut {NoStop}%
\bibitem [{\citenamefont {Matsui}\ \emph {et~al.}(2021)\citenamefont {Matsui}, \citenamefont {Katori},\ and\ \citenamefont {Shirai}}]{Ma21}%
  \BibitemOpen
\bibfield  {number} {  }\bibfield  {author} {\bibinfo {author} {\bibfnamefont {T.}~\bibnamefont {Matsui}}, \bibinfo {author} {\bibfnamefont {M.}~\bibnamefont {Katori}},\ and\ \bibinfo {author} {\bibfnamefont {T.}~\bibnamefont {Shirai}},\ }\bibfield  {title} {\bibinfo {title} {{Local number variances and hyperuniformity of the Heisenberg family of determinantal point processes}},\ }\href {https://doi.org/10.1088/1751-8121/abecaa} {\bibfield  {journal} {\bibinfo  {journal} {J. Phys. A Math. Theor.}\ }\textbf {\bibinfo {volume} {54}},\ \bibinfo {pages} {165201} (\bibinfo {year} {2021})}\BibitemShut {NoStop}%
\bibitem [{\citenamefont {Feynman}\ and\ \citenamefont {Cohen}(1956)}]{Fe56}%
  \BibitemOpen
  \bibfield  {author} {\bibinfo {author} {\bibfnamefont {R.~P.}\ \bibnamefont {Feynman}}\ and\ \bibinfo {author} {\bibfnamefont {M.}~\bibnamefont {Cohen}},\ }\bibfield  {title} {\bibinfo {title} {{Energy Spectrum of the Excitations in Liquid Helium}},\ }\href {https://doi.org/10.1103/PhysRev.102.1189} {\bibfield  {journal} {\bibinfo  {journal} {Phys. Rev.}\ }\textbf {\bibinfo {volume} {102}},\ \bibinfo {pages} {1189} (\bibinfo {year} {1956})}\BibitemShut {NoStop}%
\bibitem [{\citenamefont {Gerasimenko}\ \emph {et~al.}(2019)\citenamefont {Gerasimenko}, \citenamefont {Vaskivskyi}, \citenamefont {Litskevich}, \citenamefont {Ravnik}, \citenamefont {Vodeb}, \citenamefont {Diego}, \citenamefont {Kabanov},\ and\ \citenamefont {Mihailovic}}]{Ge19}%
  \BibitemOpen
  \bibfield  {author} {\bibinfo {author} {\bibfnamefont {Y.~A.}\ \bibnamefont {Gerasimenko}}, \bibinfo {author} {\bibfnamefont {I.}~\bibnamefont {Vaskivskyi}}, \bibinfo {author} {\bibfnamefont {M.}~\bibnamefont {Litskevich}}, \bibinfo {author} {\bibfnamefont {J.}~\bibnamefont {Ravnik}}, \bibinfo {author} {\bibfnamefont {J.}~\bibnamefont {Vodeb}}, \bibinfo {author} {\bibfnamefont {M.}~\bibnamefont {Diego}}, \bibinfo {author} {\bibfnamefont {V.}~\bibnamefont {Kabanov}},\ and\ \bibinfo {author} {\bibfnamefont {D.}~\bibnamefont {Mihailovic}},\ }\bibfield  {title} {\bibinfo {title} {{Quantum jamming transition to a correlated electron glass in 1T-TaS2}},\ }\href {https://doi.org/10.1038/s41563-019-0423-3} {\bibfield  {journal} {\bibinfo  {journal} {Nat. Mater.}\ }\textbf {\bibinfo {volume} {18}},\ \bibinfo {pages} {1078} (\bibinfo {year} {2019})}\BibitemShut {NoStop}%
\bibitem [{\citenamefont {Llorens}\ \emph {et~al.}(2020)\citenamefont {Llorens}, \citenamefont {Guillam{\'o}n}, \citenamefont {Serrano}, \citenamefont {C{\'o}rdoba}, \citenamefont {Ses{\'e}}, \citenamefont {De~Teresa}, \citenamefont {Ibarra}, \citenamefont {Vieira}, \citenamefont {Ortu{\~n}o},\ and\ \citenamefont {Suderow}}]{Ll20}%
  \BibitemOpen
  \bibfield  {author} {\bibinfo {author} {\bibfnamefont {J.~B.}\ \bibnamefont {Llorens}}, \bibinfo {author} {\bibfnamefont {I.}~\bibnamefont {Guillam{\'o}n}}, \bibinfo {author} {\bibfnamefont {I.~G.}\ \bibnamefont {Serrano}}, \bibinfo {author} {\bibfnamefont {R.}~\bibnamefont {C{\'o}rdoba}}, \bibinfo {author} {\bibfnamefont {J.}~\bibnamefont {Ses{\'e}}}, \bibinfo {author} {\bibfnamefont {J.}~\bibnamefont {De~Teresa}}, \bibinfo {author} {\bibfnamefont {M.~R.}\ \bibnamefont {Ibarra}}, \bibinfo {author} {\bibfnamefont {S.}~\bibnamefont {Vieira}}, \bibinfo {author} {\bibfnamefont {M.}~\bibnamefont {Ortu{\~n}o}},\ and\ \bibinfo {author} {\bibfnamefont {H.}~\bibnamefont {Suderow}},\ }\bibfield  {title} {\bibinfo {title} {Disordered hyperuniformity in superconducting vortex lattices},\ }\href {https://doi.org/10.1103/PhysRevResearch.2.033133} {\bibfield  {journal} {\bibinfo  {journal} {Phys. Rev. Res.}\ }\textbf {\bibinfo {volume} {2}},\ \bibinfo {pages} {033133} (\bibinfo {year} {2020})}\BibitemShut {NoStop}%
\bibitem [{\citenamefont {Sakai}\ \emph {et~al.}(2022{\natexlab{a}})\citenamefont {Sakai}, \citenamefont {Arita},\ and\ \citenamefont {Ohtsuki}}]{Sa22a}%
  \BibitemOpen
  \bibfield  {author} {\bibinfo {author} {\bibfnamefont {S.}~\bibnamefont {Sakai}}, \bibinfo {author} {\bibfnamefont {R.}~\bibnamefont {Arita}},\ and\ \bibinfo {author} {\bibfnamefont {T.}~\bibnamefont {Ohtsuki}},\ }\bibfield  {title} {\bibinfo {title} {Hyperuniform electron distributions controlled by electron interactions in quasicrystals},\ }\href {https://doi.org/10.1103/PhysRevB.105.054202} {\bibfield  {journal} {\bibinfo  {journal} {Phys. Rev. B}\ }\textbf {\bibinfo {volume} {105}},\ \bibinfo {pages} {054202} (\bibinfo {year} {2022}{\natexlab{a}})}\BibitemShut {NoStop}%
\bibitem [{\citenamefont {Sakai}\ \emph {et~al.}(2022{\natexlab{b}})\citenamefont {Sakai}, \citenamefont {Arita},\ and\ \citenamefont {Ohtsuki}}]{Sa22b}%
  \BibitemOpen
  \bibfield  {author} {\bibinfo {author} {\bibfnamefont {S.}~\bibnamefont {Sakai}}, \bibinfo {author} {\bibfnamefont {R.}~\bibnamefont {Arita}},\ and\ \bibinfo {author} {\bibfnamefont {T.}~\bibnamefont {Ohtsuki}},\ }\bibfield  {title} {\bibinfo {title} {Quantum phase transition between hyperuniform density distributions},\ }\href {https://doi.org/10.1103/PhysRevResearch.4.033241} {\bibfield  {journal} {\bibinfo  {journal} {Phys. Rev. Res.}\ }\textbf {\bibinfo {volume} {4}},\ \bibinfo {pages} {033241} (\bibinfo {year} {2022}{\natexlab{b}})}\BibitemShut {NoStop}%
\bibitem [{\citenamefont {Wang}\ \emph {et~al.}(2024)\citenamefont {Wang}, \citenamefont {Samajdar},\ and\ \citenamefont {Torquato}}]{Wa24}%
  \BibitemOpen
  \bibfield  {author} {\bibinfo {author} {\bibfnamefont {H.}~\bibnamefont {Wang}}, \bibinfo {author} {\bibfnamefont {R.}~\bibnamefont {Samajdar}},\ and\ \bibinfo {author} {\bibfnamefont {S.}~\bibnamefont {Torquato}},\ }\bibfield  {title} {\bibinfo {title} {Correlations in interacting electron liquids: Many-body statistics and hyperuniformity},\ }\href {https://doi.org/https://doi.org/10.1103/PhysRevB.110.104201} {\bibfield  {journal} {\bibinfo  {journal} {Phys. Rev. B}\ }\textbf {\bibinfo {volume} {110}},\ \bibinfo {pages} {104201} (\bibinfo {year} {2024})}\BibitemShut {NoStop}%
\bibitem [{\citenamefont {Crowley}\ \emph {et~al.}(2019)\citenamefont {Crowley}, \citenamefont {Laumann},\ and\ \citenamefont {Gopalakrishnan}}]{Cr19}%
  \BibitemOpen
  \bibfield  {author} {\bibinfo {author} {\bibfnamefont {P.~J.~D.}\ \bibnamefont {Crowley}}, \bibinfo {author} {\bibfnamefont {C.~R.}\ \bibnamefont {Laumann}},\ and\ \bibinfo {author} {\bibfnamefont {S.}~\bibnamefont {Gopalakrishnan}},\ }\bibfield  {title} {\bibinfo {title} {{Quantum criticality in Ising chains with random hyperuniform couplings}},\ }\href {https://doi.org/10.1103/PhysRevB.100.134206} {\bibfield  {journal} {\bibinfo  {journal} {Phys. Rev. B}\ }\textbf {\bibinfo {volume} {100}},\ \bibinfo {pages} {134206} (\bibinfo {year} {2019})}\BibitemShut {NoStop}%
\bibitem [{\citenamefont {Bose}\ and\ \citenamefont {Torquato}(2021)}]{Bo21}%
  \BibitemOpen
  \bibfield  {author} {\bibinfo {author} {\bibfnamefont {A.}~\bibnamefont {Bose}}\ and\ \bibinfo {author} {\bibfnamefont {S.}~\bibnamefont {Torquato}},\ }\bibfield  {title} {\bibinfo {title} {Quantum phase transitions in long-range interacting hyperuniform spin chains in a transverse field},\ }\href {https://doi.org/10.1103/PhysRevB.103.014118} {\bibfield  {journal} {\bibinfo  {journal} {Phys. Rev. B}\ }\textbf {\bibinfo {volume} {103}},\ \bibinfo {pages} {014118} (\bibinfo {year} {2021})}\BibitemShut {NoStop}%
\bibitem [{\citenamefont {Yan}\ \emph {et~al.}(2024)\citenamefont {Yan}, \citenamefont {Benton}, \citenamefont {Moessner},\ and\ \citenamefont {Nevidomskyy}}]{Ya23}%
  \BibitemOpen
  \bibfield  {author} {\bibinfo {author} {\bibfnamefont {H.}~\bibnamefont {Yan}}, \bibinfo {author} {\bibfnamefont {O.}~\bibnamefont {Benton}}, \bibinfo {author} {\bibfnamefont {R.}~\bibnamefont {Moessner}},\ and\ \bibinfo {author} {\bibfnamefont {A.~H.}\ \bibnamefont {Nevidomskyy}},\ }\bibfield  {title} {\bibinfo {title} {Classification of classical spin liquids: Typology and resulting landscape},\ }\href@noop {} {\bibfield  {journal} {\bibinfo  {journal} {Phys. Rev. B}\ }\textbf {\bibinfo {volume} {110}},\ \bibinfo {pages} {L020402} (\bibinfo {year} {2024})}\BibitemShut {NoStop}%
\bibitem [{\citenamefont {Yan}\ \emph {et~al.}(2020)\citenamefont {Yan}, \citenamefont {Benton}, \citenamefont {Jaubert},\ and\ \citenamefont {Shannon}}]{Ya20}%
  \BibitemOpen
  \bibfield  {author} {\bibinfo {author} {\bibfnamefont {H.}~\bibnamefont {Yan}}, \bibinfo {author} {\bibfnamefont {O.}~\bibnamefont {Benton}}, \bibinfo {author} {\bibfnamefont {L.~D.~C.}\ \bibnamefont {Jaubert}},\ and\ \bibinfo {author} {\bibfnamefont {N.}~\bibnamefont {Shannon}},\ }\bibfield  {title} {\bibinfo {title} {{Rank--2 $U(1)$ Spin Liquid on the Breathing Pyrochlore Lattice}},\ }\href {https://doi.org/10.1103/PhysRevLett.124.127203} {\bibfield  {journal} {\bibinfo  {journal} {Phys. Rev. Lett.}\ }\textbf {\bibinfo {volume} {124}},\ \bibinfo {pages} {127203} (\bibinfo {year} {2020})}\BibitemShut {NoStop}%
\bibitem [{\citenamefont {Benton}\ and\ \citenamefont {Moessner}(2021)}]{Be21}%
  \BibitemOpen
  \bibfield  {author} {\bibinfo {author} {\bibfnamefont {O.}~\bibnamefont {Benton}}\ and\ \bibinfo {author} {\bibfnamefont {R.}~\bibnamefont {Moessner}},\ }\bibfield  {title} {\bibinfo {title} {{Topological Route to New and Unusual Coulomb Spin Liquids}},\ }\href {https://doi.org/10.1103/PhysRevLett.127.107202} {\bibfield  {journal} {\bibinfo  {journal} {Phys. Rev. Lett.}\ }\textbf {\bibinfo {volume} {127}},\ \bibinfo {pages} {107202} (\bibinfo {year} {2021})}\BibitemShut {NoStop}%
\bibitem [{\citenamefont {Savary}\ and\ \citenamefont {Balents}(2016)}]{savary2016quantum}%
  \BibitemOpen
  \bibfield  {author} {\bibinfo {author} {\bibfnamefont {L.}~\bibnamefont {Savary}}\ and\ \bibinfo {author} {\bibfnamefont {L.}~\bibnamefont {Balents}},\ }\bibfield  {title} {\bibinfo {title} {Quantum spin liquids: a review},\ }\href {https://doi.org/10.1088/0034-4885/80/1/016502} {\bibfield  {journal} {\bibinfo  {journal} {Rep. Prog. Phys.}\ }\textbf {\bibinfo {volume} {80}},\ \bibinfo {pages} {016502} (\bibinfo {year} {2016})}\BibitemShut {NoStop}%
\bibitem [{\citenamefont {Knolle}\ and\ \citenamefont {Moessner}(2019)}]{knolle2019field}%
  \BibitemOpen
  \bibfield  {author} {\bibinfo {author} {\bibfnamefont {J.}~\bibnamefont {Knolle}}\ and\ \bibinfo {author} {\bibfnamefont {R.}~\bibnamefont {Moessner}},\ }\bibfield  {title} {\bibinfo {title} {A field guide to spin liquids},\ }\href {https://doi.org/10.1146/annurev-conmatphys-031218-013401} {\bibfield  {journal} {\bibinfo  {journal} {Annu. Rev. Condens. Matter Phys.}\ }\textbf {\bibinfo {volume} {10}},\ \bibinfo {pages} {451} (\bibinfo {year} {2019})}\BibitemShut {NoStop}%
\bibitem [{\citenamefont {Broholm}\ \emph {et~al.}(2020)\citenamefont {Broholm}, \citenamefont {Cava}, \citenamefont {Kivelson}, \citenamefont {Nocera}, \citenamefont {Norman},\ and\ \citenamefont {Senthil}}]{broholm2020quantum}%
  \BibitemOpen
  \bibfield  {author} {\bibinfo {author} {\bibfnamefont {C.}~\bibnamefont {Broholm}}, \bibinfo {author} {\bibfnamefont {R.~J.}\ \bibnamefont {Cava}}, \bibinfo {author} {\bibfnamefont {S.~A.}\ \bibnamefont {Kivelson}}, \bibinfo {author} {\bibfnamefont {D.~G.}\ \bibnamefont {Nocera}}, \bibinfo {author} {\bibfnamefont {M.~R.}\ \bibnamefont {Norman}},\ and\ \bibinfo {author} {\bibfnamefont {T.}~\bibnamefont {Senthil}},\ }\bibfield  {title} {\bibinfo {title} {Quantum spin liquids},\ }\href {https://doi.org/10.1126/science.aay0668} {\bibfield  {journal} {\bibinfo  {journal} {Science}\ }\textbf {\bibinfo {volume} {367}},\ \bibinfo {pages} {eaay0668} (\bibinfo {year} {2020})}\BibitemShut {NoStop}%
\bibitem [{\citenamefont {Read}\ and\ \citenamefont {Sachdev}(1991)}]{ReadSachdev91}%
  \BibitemOpen
  \bibfield  {author} {\bibinfo {author} {\bibfnamefont {N.}~\bibnamefont {Read}}\ and\ \bibinfo {author} {\bibfnamefont {S.}~\bibnamefont {Sachdev}},\ }\bibfield  {title} {\bibinfo {title} {{Large-N expansion for frustrated quantum antiferromagnets}},\ }\href {https://doi.org/10.1103/PhysRevLett.66.1773} {\bibfield  {journal} {\bibinfo  {journal} {Phys. Rev. Lett.}\ }\textbf {\bibinfo {volume} {66}},\ \bibinfo {pages} {1773} (\bibinfo {year} {1991})}\BibitemShut {NoStop}%
\bibitem [{\citenamefont {Wen}(1991)}]{Wen91}%
  \BibitemOpen
  \bibfield  {author} {\bibinfo {author} {\bibfnamefont {X.~G.}\ \bibnamefont {Wen}},\ }\bibfield  {title} {\bibinfo {title} {Mean-field theory of spin-liquid states with finite energy gap and topological orders},\ }\href {https://doi.org/10.1103/PhysRevB.44.2664} {\bibfield  {journal} {\bibinfo  {journal} {Phys. Rev. B}\ }\textbf {\bibinfo {volume} {44}},\ \bibinfo {pages} {2664} (\bibinfo {year} {1991})}\BibitemShut {NoStop}%
\bibitem [{\citenamefont {Sachdev}(1992)}]{Sachdev92}%
  \BibitemOpen
  \bibfield  {author} {\bibinfo {author} {\bibfnamefont {S.}~\bibnamefont {Sachdev}},\ }\bibfield  {title} {\bibinfo {title} {{Kagom{\'e}- and triangular-lattice Heisenberg antiferromagnets: Ordering from quantum fluctuations and quantum-disordered ground states with unconfined bosonic spinons}},\ }\href {https://doi.org/10.1103/PhysRevB.45.12377} {\bibfield  {journal} {\bibinfo  {journal} {Phys. Rev. B}\ }\textbf {\bibinfo {volume} {45}},\ \bibinfo {pages} {12377} (\bibinfo {year} {1992})}\BibitemShut {NoStop}%
\bibitem [{\citenamefont {Kitaev}(2006)}]{kitaev2006anyons}%
  \BibitemOpen
  \bibfield  {author} {\bibinfo {author} {\bibfnamefont {A.}~\bibnamefont {Kitaev}},\ }\bibfield  {title} {\bibinfo {title} {Anyons in an exactly solved model and beyond},\ }\href {https://doi.org/10.1016/j.aop.2005.10.005} {\bibfield  {journal} {\bibinfo  {journal} {Ann. Phys.}\ }\textbf {\bibinfo {volume} {321}},\ \bibinfo {pages} {2} (\bibinfo {year} {2006})}\BibitemShut {NoStop}%
\bibitem [{\citenamefont {Lee}(2008)}]{lee2008end}%
  \BibitemOpen
  \bibfield  {author} {\bibinfo {author} {\bibfnamefont {P.~A.}\ \bibnamefont {Lee}},\ }\bibfield  {title} {\bibinfo {title} {An end to the drought of quantum spin liquids},\ }\href {https://doi.org/10.1126/science.1163196} {\bibfield  {journal} {\bibinfo  {journal} {Science}\ }\textbf {\bibinfo {volume} {321}},\ \bibinfo {pages} {1306} (\bibinfo {year} {2008})}\BibitemShut {NoStop}%
\bibitem [{\citenamefont {Semeghini}\ \emph {et~al.}(2021)\citenamefont {Semeghini}, \citenamefont {Levine}, \citenamefont {Keesling}, \citenamefont {Ebadi}, \citenamefont {Wang}, \citenamefont {Bluvstein}, \citenamefont {Verresen}, \citenamefont {Pichler}, \citenamefont {Kalinowski}, \citenamefont {Samajdar}, \citenamefont {Omran}, \citenamefont {Sachdev}, \citenamefont {Vishwanath}, \citenamefont {Greiner}, \citenamefont {Vuleti\'c},\ and\ \citenamefont {Lukin}}]{Semeghini.2021}%
  \BibitemOpen
  \bibfield  {author} {\bibinfo {author} {\bibfnamefont {G.}~\bibnamefont {Semeghini}}, \bibinfo {author} {\bibfnamefont {H.}~\bibnamefont {Levine}}, \bibinfo {author} {\bibfnamefont {A.}~\bibnamefont {Keesling}}, \bibinfo {author} {\bibfnamefont {S.}~\bibnamefont {Ebadi}}, \bibinfo {author} {\bibfnamefont {T.~T.}\ \bibnamefont {Wang}}, \bibinfo {author} {\bibfnamefont {D.}~\bibnamefont {Bluvstein}}, \bibinfo {author} {\bibfnamefont {R.}~\bibnamefont {Verresen}}, \bibinfo {author} {\bibfnamefont {H.}~\bibnamefont {Pichler}}, \bibinfo {author} {\bibfnamefont {M.}~\bibnamefont {Kalinowski}}, \bibinfo {author} {\bibfnamefont {R.}~\bibnamefont {Samajdar}}, \bibinfo {author} {\bibfnamefont {A.}~\bibnamefont {Omran}}, \bibinfo {author} {\bibfnamefont {S.}~\bibnamefont {Sachdev}}, \bibinfo {author} {\bibfnamefont {A.}~\bibnamefont {Vishwanath}}, \bibinfo {author} {\bibfnamefont {M.}~\bibnamefont {Greiner}}, \bibinfo {author} {\bibfnamefont {V.}~\bibnamefont {Vuleti\'c}},\ and\ \bibinfo {author} {\bibfnamefont
  {M.~D.}\ \bibnamefont {Lukin}},\ }\bibfield  {title} {\bibinfo {title} {Probing topological spin liquids on a programmable quantum simulator},\ }\href {https://doi.org/10.1126/science.abi8794} {\bibfield  {journal} {\bibinfo  {journal} {Science}\ }\textbf {\bibinfo {volume} {374}},\ \bibinfo {pages} {1242} (\bibinfo {year} {2021})}\BibitemShut {NoStop}%
\bibitem [{\citenamefont {Jaksch}\ \emph {et~al.}(2000)\citenamefont {Jaksch}, \citenamefont {Cirac}, \citenamefont {Zoller}, \citenamefont {Rolston}, \citenamefont {C{\^o}t{\'e}},\ and\ \citenamefont {Lukin}}]{jaksch2000fast}%
  \BibitemOpen
  \bibfield  {author} {\bibinfo {author} {\bibfnamefont {D.}~\bibnamefont {Jaksch}}, \bibinfo {author} {\bibfnamefont {J.~I.}\ \bibnamefont {Cirac}}, \bibinfo {author} {\bibfnamefont {P.}~\bibnamefont {Zoller}}, \bibinfo {author} {\bibfnamefont {S.~L.}\ \bibnamefont {Rolston}}, \bibinfo {author} {\bibfnamefont {R.}~\bibnamefont {C{\^o}t{\'e}}},\ and\ \bibinfo {author} {\bibfnamefont {M.~D.}\ \bibnamefont {Lukin}},\ }\bibfield  {title} {\bibinfo {title} {Fast quantum gates for neutral atoms},\ }\href {https://doi.org/10.1103/PhysRevLett.85.2208} {\bibfield  {journal} {\bibinfo  {journal} {Phys. Rev. Lett.}\ }\textbf {\bibinfo {volume} {85}},\ \bibinfo {pages} {2208} (\bibinfo {year} {2000})}\BibitemShut {NoStop}%
\bibitem [{\citenamefont {Bernien}\ \emph {et~al.}(2017)\citenamefont {Bernien}, \citenamefont {Schwartz}, \citenamefont {Keesling}, \citenamefont {Levine}, \citenamefont {Omran}, \citenamefont {Pichler}, \citenamefont {Choi}, \citenamefont {Zibrov}, \citenamefont {Endres}, \citenamefont {Greiner}, \citenamefont {Vuleti{\'c}},\ and\ \citenamefont {Lukin}}]{bernien2017probing}%
  \BibitemOpen
  \bibfield  {author} {\bibinfo {author} {\bibfnamefont {H.}~\bibnamefont {Bernien}}, \bibinfo {author} {\bibfnamefont {S.}~\bibnamefont {Schwartz}}, \bibinfo {author} {\bibfnamefont {A.}~\bibnamefont {Keesling}}, \bibinfo {author} {\bibfnamefont {H.}~\bibnamefont {Levine}}, \bibinfo {author} {\bibfnamefont {A.}~\bibnamefont {Omran}}, \bibinfo {author} {\bibfnamefont {H.}~\bibnamefont {Pichler}}, \bibinfo {author} {\bibfnamefont {S.}~\bibnamefont {Choi}}, \bibinfo {author} {\bibfnamefont {A.~S.}\ \bibnamefont {Zibrov}}, \bibinfo {author} {\bibfnamefont {M.}~\bibnamefont {Endres}}, \bibinfo {author} {\bibfnamefont {M.}~\bibnamefont {Greiner}}, \bibinfo {author} {\bibfnamefont {V.}~\bibnamefont {Vuleti{\'c}}},\ and\ \bibinfo {author} {\bibfnamefont {M.~D.}\ \bibnamefont {Lukin}},\ }\bibfield  {title} {\bibinfo {title} {Probing many-body dynamics on a 51-atom quantum simulator},\ }\href {https://doi.org/10.1038/nature24622} {\bibfield  {journal} {\bibinfo  {journal} {Nature}\ }\textbf {\bibinfo {volume} {551}},\
  \bibinfo {pages} {579} (\bibinfo {year} {2017})}\BibitemShut {NoStop}%
\bibitem [{\citenamefont {de~L{\'e}s{\'e}leuc}\ \emph {et~al.}(2019)\citenamefont {de~L{\'e}s{\'e}leuc}, \citenamefont {Lienhard}, \citenamefont {Scholl}, \citenamefont {Barredo}, \citenamefont {Weber}, \citenamefont {Lang}, \citenamefont {B{\"u}chler}, \citenamefont {Lahaye},\ and\ \citenamefont {Browaeys}}]{de2019observation}%
  \BibitemOpen
  \bibfield  {author} {\bibinfo {author} {\bibfnamefont {S.}~\bibnamefont {de~L{\'e}s{\'e}leuc}}, \bibinfo {author} {\bibfnamefont {V.}~\bibnamefont {Lienhard}}, \bibinfo {author} {\bibfnamefont {P.}~\bibnamefont {Scholl}}, \bibinfo {author} {\bibfnamefont {D.}~\bibnamefont {Barredo}}, \bibinfo {author} {\bibfnamefont {S.}~\bibnamefont {Weber}}, \bibinfo {author} {\bibfnamefont {N.}~\bibnamefont {Lang}}, \bibinfo {author} {\bibfnamefont {H.~P.}\ \bibnamefont {B{\"u}chler}}, \bibinfo {author} {\bibfnamefont {T.}~\bibnamefont {Lahaye}},\ and\ \bibinfo {author} {\bibfnamefont {A.}~\bibnamefont {Browaeys}},\ }\bibfield  {title} {\bibinfo {title} {{Observation of a symmetry-protected topological phase of interacting bosons with Rydberg atoms}},\ }\href {https://doi.org/10.1126/science.aav9105} {\bibfield  {journal} {\bibinfo  {journal} {Science}\ }\textbf {\bibinfo {volume} {365}},\ \bibinfo {pages} {775} (\bibinfo {year} {2019})}\BibitemShut {NoStop}%
\bibitem [{\citenamefont {Samajdar}\ \emph {et~al.}(2020)\citenamefont {Samajdar}, \citenamefont {Ho}, \citenamefont {Pichler}, \citenamefont {Lukin},\ and\ \citenamefont {Sachdev}}]{Samajdar_2020}%
  \BibitemOpen
  \bibfield  {author} {\bibinfo {author} {\bibfnamefont {R.}~\bibnamefont {Samajdar}}, \bibinfo {author} {\bibfnamefont {W.~W.}\ \bibnamefont {Ho}}, \bibinfo {author} {\bibfnamefont {H.}~\bibnamefont {Pichler}}, \bibinfo {author} {\bibfnamefont {M.~D.}\ \bibnamefont {Lukin}},\ and\ \bibinfo {author} {\bibfnamefont {S.}~\bibnamefont {Sachdev}},\ }\bibfield  {title} {\bibinfo {title} {{Complex Density Wave Orders and Quantum Phase Transitions in a Model of Square-Lattice Rydberg Atom Arrays}},\ }\href {https://doi.org/10.1103/physrevlett.124.103601} {\bibfield  {journal} {\bibinfo  {journal} {Phys. Rev. Lett.}\ }\textbf {\bibinfo {volume} {124}},\ \bibinfo {pages} {103601} (\bibinfo {year} {2020})}\BibitemShut {NoStop}%
\bibitem [{\citenamefont {Felser}\ \emph {et~al.}(2021)\citenamefont {Felser}, \citenamefont {Notarnicola},\ and\ \citenamefont {Montangero}}]{PhysRevLett.126.170603}%
  \BibitemOpen
  \bibfield  {author} {\bibinfo {author} {\bibfnamefont {T.}~\bibnamefont {Felser}}, \bibinfo {author} {\bibfnamefont {S.}~\bibnamefont {Notarnicola}},\ and\ \bibinfo {author} {\bibfnamefont {S.}~\bibnamefont {Montangero}},\ }\bibfield  {title} {\bibinfo {title} {{Efficient Tensor Network Ansatz for High-Dimensional Quantum Many-Body Problems}},\ }\href {https://doi.org/10.1103/PhysRevLett.126.170603} {\bibfield  {journal} {\bibinfo  {journal} {Phys. Rev. Lett.}\ }\textbf {\bibinfo {volume} {126}},\ \bibinfo {pages} {170603} (\bibinfo {year} {2021})}\BibitemShut {NoStop}%
\bibitem [{\citenamefont {Ebadi}\ \emph {et~al.}(2021)\citenamefont {Ebadi}, \citenamefont {Wang}, \citenamefont {Levine}, \citenamefont {Keesling}, \citenamefont {Semeghini}, \citenamefont {Omran}, \citenamefont {Bluvstein}, \citenamefont {Samajdar}, \citenamefont {Pichler}, \citenamefont {Ho}, \citenamefont {Choi}, \citenamefont {Sachdev}, \citenamefont {Greiner}, \citenamefont {Vuleti\'c},\ and\ \citenamefont {Lukin}}]{Ebadi.2021}%
  \BibitemOpen
  \bibfield  {author} {\bibinfo {author} {\bibfnamefont {S.}~\bibnamefont {Ebadi}}, \bibinfo {author} {\bibfnamefont {T.~T.}\ \bibnamefont {Wang}}, \bibinfo {author} {\bibfnamefont {H.}~\bibnamefont {Levine}}, \bibinfo {author} {\bibfnamefont {A.}~\bibnamefont {Keesling}}, \bibinfo {author} {\bibfnamefont {G.}~\bibnamefont {Semeghini}}, \bibinfo {author} {\bibfnamefont {A.}~\bibnamefont {Omran}}, \bibinfo {author} {\bibfnamefont {D.}~\bibnamefont {Bluvstein}}, \bibinfo {author} {\bibfnamefont {R.}~\bibnamefont {Samajdar}}, \bibinfo {author} {\bibfnamefont {H.}~\bibnamefont {Pichler}}, \bibinfo {author} {\bibfnamefont {W.~W.}\ \bibnamefont {Ho}}, \bibinfo {author} {\bibfnamefont {S.}~\bibnamefont {Choi}}, \bibinfo {author} {\bibfnamefont {S.}~\bibnamefont {Sachdev}}, \bibinfo {author} {\bibfnamefont {M.}~\bibnamefont {Greiner}}, \bibinfo {author} {\bibfnamefont {V.}~\bibnamefont {Vuleti\'c}},\ and\ \bibinfo {author} {\bibfnamefont {M.~D.}\ \bibnamefont {Lukin}},\ }\bibfield  {title} {\bibinfo {title} {{Quantum
  phases of matter on a 256-atom programmable quantum simulator}},\ }\href {https://doi.org/10.1038/s41586-021-03582-4} {\bibfield  {journal} {\bibinfo  {journal} {Nature}\ }\textbf {\bibinfo {volume} {595}},\ \bibinfo {pages} {227} (\bibinfo {year} {2021})}\BibitemShut {NoStop}%
\bibitem [{\citenamefont {Scholl}\ \emph {et~al.}(2021)\citenamefont {Scholl}, \citenamefont {Schuler}, \citenamefont {Williams}, \citenamefont {Eberharter}, \citenamefont {Barredo}, \citenamefont {Schymik}, \citenamefont {Lienhard}, \citenamefont {Henry}, \citenamefont {Lang}, \citenamefont {Lahaye}, \citenamefont {L\"auchli},\ and\ \citenamefont {Browaeys}}]{scholl2021quantum}%
  \BibitemOpen
  \bibfield  {author} {\bibinfo {author} {\bibfnamefont {P.}~\bibnamefont {Scholl}}, \bibinfo {author} {\bibfnamefont {M.}~\bibnamefont {Schuler}}, \bibinfo {author} {\bibfnamefont {H.~J.}\ \bibnamefont {Williams}}, \bibinfo {author} {\bibfnamefont {A.~A.}\ \bibnamefont {Eberharter}}, \bibinfo {author} {\bibfnamefont {D.}~\bibnamefont {Barredo}}, \bibinfo {author} {\bibfnamefont {K.-N.}\ \bibnamefont {Schymik}}, \bibinfo {author} {\bibfnamefont {V.}~\bibnamefont {Lienhard}}, \bibinfo {author} {\bibfnamefont {L.-P.}\ \bibnamefont {Henry}}, \bibinfo {author} {\bibfnamefont {T.~C.}\ \bibnamefont {Lang}}, \bibinfo {author} {\bibfnamefont {T.}~\bibnamefont {Lahaye}}, \bibinfo {author} {\bibfnamefont {A.~M.}\ \bibnamefont {L\"auchli}},\ and\ \bibinfo {author} {\bibfnamefont {A.}~\bibnamefont {Browaeys}},\ }\bibfield  {title} {\bibinfo {title} {{Quantum simulation of 2D antiferromagnets with hundreds of Rydberg atoms}},\ }\href {https://doi.org/10.1038/s41586-021-03585-1} {\bibfield  {journal} {\bibinfo  {journal}
  {Nature}\ }\textbf {\bibinfo {volume} {595}},\ \bibinfo {pages} {233} (\bibinfo {year} {2021})}\BibitemShut {NoStop}%
\bibitem [{\citenamefont {Miles}\ \emph {et~al.}(2023)\citenamefont {Miles}, \citenamefont {Samajdar}, \citenamefont {Ebadi}, \citenamefont {Wang}, \citenamefont {Pichler}, \citenamefont {Sachdev}, \citenamefont {Lukin}, \citenamefont {Greiner}, \citenamefont {Weinberger},\ and\ \citenamefont {Kim}}]{Kim.2021}%
  \BibitemOpen
  \bibfield  {author} {\bibinfo {author} {\bibfnamefont {C.}~\bibnamefont {Miles}}, \bibinfo {author} {\bibfnamefont {R.}~\bibnamefont {Samajdar}}, \bibinfo {author} {\bibfnamefont {S.}~\bibnamefont {Ebadi}}, \bibinfo {author} {\bibfnamefont {T.~T.}\ \bibnamefont {Wang}}, \bibinfo {author} {\bibfnamefont {H.}~\bibnamefont {Pichler}}, \bibinfo {author} {\bibfnamefont {S.}~\bibnamefont {Sachdev}}, \bibinfo {author} {\bibfnamefont {M.~D.}\ \bibnamefont {Lukin}}, \bibinfo {author} {\bibfnamefont {M.}~\bibnamefont {Greiner}}, \bibinfo {author} {\bibfnamefont {K.~Q.}\ \bibnamefont {Weinberger}},\ and\ \bibinfo {author} {\bibfnamefont {E.-A.}\ \bibnamefont {Kim}},\ }\bibfield  {title} {\bibinfo {title} {Machine learning discovery of new phases in programmable quantum simulator snapshots},\ }\href {https://doi.org/10.1103/PhysRevResearch.5.013026} {\bibfield  {journal} {\bibinfo  {journal} {Phys. Rev. Res.}\ }\textbf {\bibinfo {volume} {5}},\ \bibinfo {pages} {013026} (\bibinfo {year} {2023})}\BibitemShut {NoStop}%
\bibitem [{\citenamefont {Samajdar}\ \emph {et~al.}(2018)\citenamefont {Samajdar}, \citenamefont {Choi}, \citenamefont {Pichler}, \citenamefont {Lukin},\ and\ \citenamefont {Sachdev}}]{samajdar2018numerical}%
  \BibitemOpen
  \bibfield  {author} {\bibinfo {author} {\bibfnamefont {R.}~\bibnamefont {Samajdar}}, \bibinfo {author} {\bibfnamefont {S.}~\bibnamefont {Choi}}, \bibinfo {author} {\bibfnamefont {H.}~\bibnamefont {Pichler}}, \bibinfo {author} {\bibfnamefont {M.~D.}\ \bibnamefont {Lukin}},\ and\ \bibinfo {author} {\bibfnamefont {S.}~\bibnamefont {Sachdev}},\ }\bibfield  {title} {\bibinfo {title} {Numerical study of the chiral $\mathbb{Z}_3$ quantum phase transition in one spatial dimension},\ }\href {https://doi.org/10.1103/PhysRevA.98.023614} {\bibfield  {journal} {\bibinfo  {journal} {Phys. Rev. A}\ }\textbf {\bibinfo {volume} {98}},\ \bibinfo {pages} {023614} (\bibinfo {year} {2018})}\BibitemShut {NoStop}%
\bibitem [{\citenamefont {Whitsitt}\ \emph {et~al.}(2018)\citenamefont {Whitsitt}, \citenamefont {Samajdar},\ and\ \citenamefont {Sachdev}}]{whitsitt2018quantum}%
  \BibitemOpen
  \bibfield  {author} {\bibinfo {author} {\bibfnamefont {S.}~\bibnamefont {Whitsitt}}, \bibinfo {author} {\bibfnamefont {R.}~\bibnamefont {Samajdar}},\ and\ \bibinfo {author} {\bibfnamefont {S.}~\bibnamefont {Sachdev}},\ }\bibfield  {title} {\bibinfo {title} {Quantum field theory for the chiral clock transition in one spatial dimension},\ }\href {https://doi.org/10.1103/PhysRevB.98.205118} {\bibfield  {journal} {\bibinfo  {journal} {Phys. Rev. B}\ }\textbf {\bibinfo {volume} {98}},\ \bibinfo {pages} {205118} (\bibinfo {year} {2018})}\BibitemShut {NoStop}%
\bibitem [{\citenamefont {Chepiga}\ and\ \citenamefont {Mila}(2019)}]{PhysRevLett.122.017205}%
  \BibitemOpen
  \bibfield  {author} {\bibinfo {author} {\bibfnamefont {N.}~\bibnamefont {Chepiga}}\ and\ \bibinfo {author} {\bibfnamefont {F.}~\bibnamefont {Mila}},\ }\bibfield  {title} {\bibinfo {title} {{Floating Phase versus Chiral Transition in a 1D Hard-Boson Model}},\ }\href {https://doi.org/10.1103/PhysRevLett.122.017205} {\bibfield  {journal} {\bibinfo  {journal} {Phys. Rev. Lett.}\ }\textbf {\bibinfo {volume} {122}},\ \bibinfo {pages} {017205} (\bibinfo {year} {2019})}\BibitemShut {NoStop}%
\bibitem [{\citenamefont {Chepiga}\ and\ \citenamefont {Mila}(2021)}]{chepiga2021kibble}%
  \BibitemOpen
  \bibfield  {author} {\bibinfo {author} {\bibfnamefont {N.}~\bibnamefont {Chepiga}}\ and\ \bibinfo {author} {\bibfnamefont {F.}~\bibnamefont {Mila}},\ }\bibfield  {title} {\bibinfo {title} {{Kibble-Zurek exponent and chiral transition of the period-4 phase of Rydberg chains}},\ }\href {https://doi.org/10.1038/s41467-020-20641-y} {\bibfield  {journal} {\bibinfo  {journal} {Nat. Commun.}\ }\textbf {\bibinfo {volume} {12}},\ \bibinfo {pages} {1} (\bibinfo {year} {2021})}\BibitemShut {NoStop}%
\bibitem [{\citenamefont {Kalinowski}\ \emph {et~al.}(2022)\citenamefont {Kalinowski}, \citenamefont {Samajdar}, \citenamefont {Melko}, \citenamefont {Lukin}, \citenamefont {Sachdev},\ and\ \citenamefont {Choi}}]{kalinowski2021bulk}%
  \BibitemOpen
  \bibfield  {author} {\bibinfo {author} {\bibfnamefont {M.}~\bibnamefont {Kalinowski}}, \bibinfo {author} {\bibfnamefont {R.}~\bibnamefont {Samajdar}}, \bibinfo {author} {\bibfnamefont {R.~G.}\ \bibnamefont {Melko}}, \bibinfo {author} {\bibfnamefont {M.~D.}\ \bibnamefont {Lukin}}, \bibinfo {author} {\bibfnamefont {S.}~\bibnamefont {Sachdev}},\ and\ \bibinfo {author} {\bibfnamefont {S.}~\bibnamefont {Choi}},\ }\bibfield  {title} {\bibinfo {title} {{Bulk and boundary quantum phase transitions in a square Rydberg atom array}},\ }\href {https://doi.org/10.1103/PhysRevB.105.174417} {\bibfield  {journal} {\bibinfo  {journal} {Phys. Rev. B}\ }\textbf {\bibinfo {volume} {105}},\ \bibinfo {pages} {174417} (\bibinfo {year} {2022})}\BibitemShut {NoStop}%
\bibitem [{\citenamefont {Turner}\ \emph {et~al.}(2018)\citenamefont {Turner}, \citenamefont {Michailidis}, \citenamefont {Abanin}, \citenamefont {Serbyn},\ and\ \citenamefont {Papi{\'c}}}]{turner2018weak}%
  \BibitemOpen
  \bibfield  {author} {\bibinfo {author} {\bibfnamefont {C.~J.}\ \bibnamefont {Turner}}, \bibinfo {author} {\bibfnamefont {A.~A.}\ \bibnamefont {Michailidis}}, \bibinfo {author} {\bibfnamefont {D.~A.}\ \bibnamefont {Abanin}}, \bibinfo {author} {\bibfnamefont {M.}~\bibnamefont {Serbyn}},\ and\ \bibinfo {author} {\bibfnamefont {Z.}~\bibnamefont {Papi{\'c}}},\ }\bibfield  {title} {\bibinfo {title} {Weak ergodicity breaking from quantum many-body scars},\ }\href {https://doi.org/10.1038/s41567-018-0137-5} {\bibfield  {journal} {\bibinfo  {journal} {Nature Phys.}\ }\textbf {\bibinfo {volume} {14}},\ \bibinfo {pages} {745} (\bibinfo {year} {2018})}\BibitemShut {NoStop}%
\bibitem [{\citenamefont {Keesling}\ \emph {et~al.}(2019)\citenamefont {Keesling}, \citenamefont {Omran}, \citenamefont {Levine}, \citenamefont {Bernien}, \citenamefont {Pichler}, \citenamefont {Choi}, \citenamefont {Samajdar}, \citenamefont {Schwartz}, \citenamefont {Silvi}, \citenamefont {Sachdev}, \citenamefont {Zoller}, \citenamefont {Endres}, \citenamefont {Greiner}, \citenamefont {Vuleti{\'c}},\ and\ \citenamefont {Lukin}}]{keesling2019quantum}%
  \BibitemOpen
  \bibfield  {author} {\bibinfo {author} {\bibfnamefont {A.}~\bibnamefont {Keesling}}, \bibinfo {author} {\bibfnamefont {A.}~\bibnamefont {Omran}}, \bibinfo {author} {\bibfnamefont {H.}~\bibnamefont {Levine}}, \bibinfo {author} {\bibfnamefont {H.}~\bibnamefont {Bernien}}, \bibinfo {author} {\bibfnamefont {H.}~\bibnamefont {Pichler}}, \bibinfo {author} {\bibfnamefont {S.}~\bibnamefont {Choi}}, \bibinfo {author} {\bibfnamefont {R.}~\bibnamefont {Samajdar}}, \bibinfo {author} {\bibfnamefont {S.}~\bibnamefont {Schwartz}}, \bibinfo {author} {\bibfnamefont {P.}~\bibnamefont {Silvi}}, \bibinfo {author} {\bibfnamefont {S.}~\bibnamefont {Sachdev}}, \bibinfo {author} {\bibfnamefont {P.}~\bibnamefont {Zoller}}, \bibinfo {author} {\bibfnamefont {M.}~\bibnamefont {Endres}}, \bibinfo {author} {\bibfnamefont {M.}~\bibnamefont {Greiner}}, \bibinfo {author} {\bibfnamefont {V.}~\bibnamefont {Vuleti{\'c}}},\ and\ \bibinfo {author} {\bibfnamefont {M.~D.}\ \bibnamefont {Lukin}},\ }\bibfield  {title} {\bibinfo {title} {{Quantum
  Kibble--Zurek mechanism and critical dynamics on a programmable Rydberg simulator}},\ }\href {https://doi.org/10.1038/s41586-019-1070-1} {\bibfield  {journal} {\bibinfo  {journal} {Nature}\ }\textbf {\bibinfo {volume} {568}},\ \bibinfo {pages} {207} (\bibinfo {year} {2019})}\BibitemShut {NoStop}%
\bibitem [{\citenamefont {Bluvstein}\ \emph {et~al.}(2021)\citenamefont {Bluvstein}, \citenamefont {Omran}, \citenamefont {Levine}, \citenamefont {Keesling}, \citenamefont {Semeghini}, \citenamefont {Ebadi}, \citenamefont {Wang}, \citenamefont {Michailidis}, \citenamefont {Maskara}, \citenamefont {Ho}, \citenamefont {Choi}, \citenamefont {Serbyn}, \citenamefont {Greiner}, \citenamefont {Vuleti\'c},\ and\ \citenamefont {Lukin}}]{Bluvstein.2021}%
  \BibitemOpen
  \bibfield  {author} {\bibinfo {author} {\bibfnamefont {D.}~\bibnamefont {Bluvstein}}, \bibinfo {author} {\bibfnamefont {A.}~\bibnamefont {Omran}}, \bibinfo {author} {\bibfnamefont {H.}~\bibnamefont {Levine}}, \bibinfo {author} {\bibfnamefont {A.}~\bibnamefont {Keesling}}, \bibinfo {author} {\bibfnamefont {G.}~\bibnamefont {Semeghini}}, \bibinfo {author} {\bibfnamefont {S.}~\bibnamefont {Ebadi}}, \bibinfo {author} {\bibfnamefont {T.~T.}\ \bibnamefont {Wang}}, \bibinfo {author} {\bibfnamefont {A.~A.}\ \bibnamefont {Michailidis}}, \bibinfo {author} {\bibfnamefont {N.}~\bibnamefont {Maskara}}, \bibinfo {author} {\bibfnamefont {W.~W.}\ \bibnamefont {Ho}}, \bibinfo {author} {\bibfnamefont {S.}~\bibnamefont {Choi}}, \bibinfo {author} {\bibfnamefont {M.}~\bibnamefont {Serbyn}}, \bibinfo {author} {\bibfnamefont {M.}~\bibnamefont {Greiner}}, \bibinfo {author} {\bibfnamefont {V.}~\bibnamefont {Vuleti\'c}},\ and\ \bibinfo {author} {\bibfnamefont {M.~D.}\ \bibnamefont {Lukin}},\ }\bibfield  {title} {\bibinfo {title}
  {{Controlling quantum many-body dynamics in driven Rydberg atom arrays}},\ }\href {https://doi.org/10.1126/science.abg2530} {\bibfield  {journal} {\bibinfo  {journal} {Science}\ }\textbf {\bibinfo {volume} {371}},\ \bibinfo {pages} {1355} (\bibinfo {year} {2021})},\ \Eprint {https://arxiv.org/abs/2012.12276} {2012.12276} \BibitemShut {NoStop}%
\bibitem [{\citenamefont {Samajdar}\ \emph {et~al.}(2021)\citenamefont {Samajdar}, \citenamefont {Ho}, \citenamefont {Pichler}, \citenamefont {Lukin},\ and\ \citenamefont {Sachdev}}]{Samajdar.2021}%
  \BibitemOpen
  \bibfield  {author} {\bibinfo {author} {\bibfnamefont {R.}~\bibnamefont {Samajdar}}, \bibinfo {author} {\bibfnamefont {W.~W.}\ \bibnamefont {Ho}}, \bibinfo {author} {\bibfnamefont {H.}~\bibnamefont {Pichler}}, \bibinfo {author} {\bibfnamefont {M.~D.}\ \bibnamefont {Lukin}},\ and\ \bibinfo {author} {\bibfnamefont {S.}~\bibnamefont {Sachdev}},\ }\bibfield  {title} {\bibinfo {title} {{Quantum phases of Rydberg atoms on a kagome lattice}},\ }\href {https://doi.org/10.1073/pnas.2015785118} {\bibfield  {journal} {\bibinfo  {journal} {Proc. Natl. Acad. Sci. U.S.A.}\ }\textbf {\bibinfo {volume} {118}},\ \bibinfo {pages} {e2015785118} (\bibinfo {year} {2021})},\ \Eprint {https://arxiv.org/abs/2011.12295} {2011.12295} \BibitemShut {NoStop}%
\bibitem [{\citenamefont {Verresen}\ \emph {et~al.}(2021)\citenamefont {Verresen}, \citenamefont {Lukin},\ and\ \citenamefont {Vishwanath}}]{Verresen.2020}%
  \BibitemOpen
  \bibfield  {author} {\bibinfo {author} {\bibfnamefont {R.}~\bibnamefont {Verresen}}, \bibinfo {author} {\bibfnamefont {M.~D.}\ \bibnamefont {Lukin}},\ and\ \bibinfo {author} {\bibfnamefont {A.}~\bibnamefont {Vishwanath}},\ }\bibfield  {title} {\bibinfo {title} {{Prediction of Toric Code Topological Order from Rydberg Blockade}},\ }\href {https://doi.org/10.1103/PhysRevX.11.031005} {\bibfield  {journal} {\bibinfo  {journal} {Phys. Rev. X}\ }\textbf {\bibinfo {volume} {11}},\ \bibinfo {pages} {031005} (\bibinfo {year} {2021})}\BibitemShut {NoStop}%
\bibitem [{\citenamefont {Yan}\ \emph {et~al.}(2022)\citenamefont {Yan}, \citenamefont {Samajdar}, \citenamefont {Wang}, \citenamefont {Sachdev},\ and\ \citenamefont {Meng}}]{yan2022triangular}%
  \BibitemOpen
  \bibfield  {author} {\bibinfo {author} {\bibfnamefont {Z.}~\bibnamefont {Yan}}, \bibinfo {author} {\bibfnamefont {R.}~\bibnamefont {Samajdar}}, \bibinfo {author} {\bibfnamefont {Y.-C.}\ \bibnamefont {Wang}}, \bibinfo {author} {\bibfnamefont {S.}~\bibnamefont {Sachdev}},\ and\ \bibinfo {author} {\bibfnamefont {Z.~Y.}\ \bibnamefont {Meng}},\ }\bibfield  {title} {\bibinfo {title} {Triangular lattice quantum dimer model with variable dimer density},\ }\href {https://doi.org/10.1038/s41467-022-33431-5} {\bibfield  {journal} {\bibinfo  {journal} {Nat. Commun.}\ }\textbf {\bibinfo {volume} {13}},\ \bibinfo {pages} {5799} (\bibinfo {year} {2022})}\BibitemShut {NoStop}%
\bibitem [{\citenamefont {Moessner}\ and\ \citenamefont {Sondhi}(2001{\natexlab{a}})}]{moessner2001ising}%
  \BibitemOpen
  \bibfield  {author} {\bibinfo {author} {\bibfnamefont {R.}~\bibnamefont {Moessner}}\ and\ \bibinfo {author} {\bibfnamefont {S.~L.}\ \bibnamefont {Sondhi}},\ }\bibfield  {title} {\bibinfo {title} {Ising models of quantum frustration},\ }\href {https://doi.org/10.1103/PhysRevB.63.224401} {\bibfield  {journal} {\bibinfo  {journal} {Phys. Rev. B}\ }\textbf {\bibinfo {volume} {63}},\ \bibinfo {pages} {224401} (\bibinfo {year} {2001}{\natexlab{a}})}\BibitemShut {NoStop}%
\bibitem [{\citenamefont {Moessner}\ and\ \citenamefont {Raman}(2011)}]{moessner2008quantum}%
  \BibitemOpen
  \bibfield  {author} {\bibinfo {author} {\bibfnamefont {R.}~\bibnamefont {Moessner}}\ and\ \bibinfo {author} {\bibfnamefont {K.~S.}\ \bibnamefont {Raman}},\ }\bibfield  {title} {\bibinfo {title} {Quantum dimer models},\ }in\ \href {https://doi.org/10.1007/978-3-642-10589-0_17} {\emph {\bibinfo {booktitle} {{Introduction to Frustrated Magnetism}}}}\ (\bibinfo  {publisher} {Springer},\ \bibinfo {year} {2011})\ pp.\ \bibinfo {pages} {437--479}\BibitemShut {NoStop}%
\bibitem [{\citenamefont {Misguich}\ \emph {et~al.}(2002)\citenamefont {Misguich}, \citenamefont {Serban},\ and\ \citenamefont {Pasquier}}]{misguich2002quantum}%
  \BibitemOpen
  \bibfield  {author} {\bibinfo {author} {\bibfnamefont {G.}~\bibnamefont {Misguich}}, \bibinfo {author} {\bibfnamefont {D.}~\bibnamefont {Serban}},\ and\ \bibinfo {author} {\bibfnamefont {V.}~\bibnamefont {Pasquier}},\ }\bibfield  {title} {\bibinfo {title} {{Quantum Dimer Model on the Kagome Lattice: Solvable Dimer-Liquid and Ising Gauge Theory}},\ }\href {https://doi.org/10.1103/PhysRevLett.89.137202} {\bibfield  {journal} {\bibinfo  {journal} {Phys. Rev. Lett.}\ }\textbf {\bibinfo {volume} {89}},\ \bibinfo {pages} {137202} (\bibinfo {year} {2002})}\BibitemShut {NoStop}%
\bibitem [{\citenamefont {Misguich}\ \emph {et~al.}(2003)\citenamefont {Misguich}, \citenamefont {Serban},\ and\ \citenamefont {Pasquier}}]{misguich2003quantum}%
  \BibitemOpen
  \bibfield  {author} {\bibinfo {author} {\bibfnamefont {G.}~\bibnamefont {Misguich}}, \bibinfo {author} {\bibfnamefont {D.}~\bibnamefont {Serban}},\ and\ \bibinfo {author} {\bibfnamefont {V.}~\bibnamefont {Pasquier}},\ }\bibfield  {title} {\bibinfo {title} {{Quantum dimer model with extensive ground-state entropy on the kagome lattice}},\ }\href {https://doi.org/10.1103/PhysRevB.67.214413} {\bibfield  {journal} {\bibinfo  {journal} {Phys. Rev. B}\ }\textbf {\bibinfo {volume} {67}},\ \bibinfo {pages} {214413} (\bibinfo {year} {2003})}\BibitemShut {NoStop}%
\bibitem [{\citenamefont {Misguich}\ \emph {et~al.}(2004)\citenamefont {Misguich}, \citenamefont {Serban},\ and\ \citenamefont {Pasquier}}]{misguich2004interaction}%
  \BibitemOpen
  \bibfield  {author} {\bibinfo {author} {\bibfnamefont {G.}~\bibnamefont {Misguich}}, \bibinfo {author} {\bibfnamefont {D.}~\bibnamefont {Serban}},\ and\ \bibinfo {author} {\bibfnamefont {V.}~\bibnamefont {Pasquier}},\ }\bibfield  {title} {\bibinfo {title} {Interaction between static holes in a quantum dimer model on the kagome lattice},\ }\href {https://doi.org/10.1088/0953-8984/16/11/036} {\bibfield  {journal} {\bibinfo  {journal} {J. Phys.: Condens. Matter}\ }\textbf {\bibinfo {volume} {16}},\ \bibinfo {pages} {S823} (\bibinfo {year} {2004})}\BibitemShut {NoStop}%
\bibitem [{\citenamefont {Elser}\ and\ \citenamefont {Zeng}(1993)}]{elser1993kagome}%
  \BibitemOpen
  \bibfield  {author} {\bibinfo {author} {\bibfnamefont {V.}~\bibnamefont {Elser}}\ and\ \bibinfo {author} {\bibfnamefont {C.}~\bibnamefont {Zeng}},\ }\bibfield  {title} {\bibinfo {title} {{Kagome spin-1/2 antiferromagnets in the hyperbolic plane}},\ }\href {https://doi.org/10.1103/PhysRevB.48.13647} {\bibfield  {journal} {\bibinfo  {journal} {Phys. Rev. B}\ }\textbf {\bibinfo {volume} {48}},\ \bibinfo {pages} {13647} (\bibinfo {year} {1993})}\BibitemShut {NoStop}%
\bibitem [{\citenamefont {Fisher}(1966)}]{fisher1966dimer}%
  \BibitemOpen
  \bibfield  {author} {\bibinfo {author} {\bibfnamefont {M.~E.}\ \bibnamefont {Fisher}},\ }\bibfield  {title} {\bibinfo {title} {{On the dimer solution of planar Ising models}},\ }\href {https://doi.org/10.1063/1.1704825} {\bibfield  {journal} {\bibinfo  {journal} {J. Math. Phys.}\ }\textbf {\bibinfo {volume} {7}},\ \bibinfo {pages} {1776} (\bibinfo {year} {1966})}\BibitemShut {NoStop}%
\bibitem [{\citenamefont {Moessner}\ and\ \citenamefont {Sondhi}(2003)}]{mossner2003ising}%
  \BibitemOpen
  \bibfield  {author} {\bibinfo {author} {\bibfnamefont {R.}~\bibnamefont {Moessner}}\ and\ \bibinfo {author} {\bibfnamefont {S.~L.}\ \bibnamefont {Sondhi}},\ }\bibfield  {title} {\bibinfo {title} {{Ising and dimer models in two and three dimensions}},\ }\href {https://doi.org/10.1103/PhysRevB.68.054405} {\bibfield  {journal} {\bibinfo  {journal} {Phys. Rev. B}\ }\textbf {\bibinfo {volume} {68}},\ \bibinfo {pages} {054405} (\bibinfo {year} {2003})}\BibitemShut {NoStop}%
\bibitem [{\citenamefont {Balents}(2010)}]{balents2010spin}%
  \BibitemOpen
  \bibfield  {author} {\bibinfo {author} {\bibfnamefont {L.}~\bibnamefont {Balents}},\ }\bibfield  {title} {\bibinfo {title} {Spin liquids in frustrated magnets},\ }\href {https://doi.org/10.1038/nature08917} {\bibfield  {journal} {\bibinfo  {journal} {Nature}\ }\textbf {\bibinfo {volume} {464}},\ \bibinfo {pages} {199} (\bibinfo {year} {2010})}\BibitemShut {NoStop}%
\bibitem [{\citenamefont {DiStasio~Jr}\ \emph {et~al.}(2018)\citenamefont {DiStasio~Jr}, \citenamefont {Zhang}, \citenamefont {Stillinger},\ and\ \citenamefont {Torquato}}]{Di18a}%
  \BibitemOpen
  \bibfield  {author} {\bibinfo {author} {\bibfnamefont {R.~A.}\ \bibnamefont {DiStasio~Jr}}, \bibinfo {author} {\bibfnamefont {G.}~\bibnamefont {Zhang}}, \bibinfo {author} {\bibfnamefont {F.~H.}\ \bibnamefont {Stillinger}},\ and\ \bibinfo {author} {\bibfnamefont {S.}~\bibnamefont {Torquato}},\ }\bibfield  {title} {\bibinfo {title} {Rational design of stealthy hyperuniform two-phase media with tunable order},\ }\href {https://doi.org/https://doi.org/10.1103/PhysRevE.97.023311} {\bibfield  {journal} {\bibinfo  {journal} {Phys. Rev. E}\ }\textbf {\bibinfo {volume} {97}},\ \bibinfo {pages} {023311} (\bibinfo {year} {2018})}\BibitemShut {NoStop}%
\bibitem [{\citenamefont {Chen}\ \emph {et~al.}(2023)\citenamefont {Chen}, \citenamefont {Jiang}, \citenamefont {Wang}, \citenamefont {Vidallon}, \citenamefont {Zhuang},\ and\ \citenamefont {Jiao}}]{Ch23}%
  \BibitemOpen
  \bibfield  {author} {\bibinfo {author} {\bibfnamefont {D.}~\bibnamefont {Chen}}, \bibinfo {author} {\bibfnamefont {X.}~\bibnamefont {Jiang}}, \bibinfo {author} {\bibfnamefont {D.}~\bibnamefont {Wang}}, \bibinfo {author} {\bibfnamefont {J.~I.}\ \bibnamefont {Vidallon}}, \bibinfo {author} {\bibfnamefont {H.}~\bibnamefont {Zhuang}},\ and\ \bibinfo {author} {\bibfnamefont {Y.}~\bibnamefont {Jiao}},\ }\bibfield  {title} {\bibinfo {title} {Multihyperuniform long-range order in medium-entropy alloys},\ }\href {https://doi.org/https://doi.org/10.1016/j.actamat.2023.118678} {\bibfield  {journal} {\bibinfo  {journal} {Acta Mater.}\ }\textbf {\bibinfo {volume} {246}},\ \bibinfo {pages} {118678} (\bibinfo {year} {2023})}\BibitemShut {NoStop}%
\bibitem [{\citenamefont {Wang}\ and\ \citenamefont {Torquato}(2022)}]{Wa22}%
  \BibitemOpen
  \bibfield  {author} {\bibinfo {author} {\bibfnamefont {H.}~\bibnamefont {Wang}}\ and\ \bibinfo {author} {\bibfnamefont {S.}~\bibnamefont {Torquato}},\ }\bibfield  {title} {\bibinfo {title} {Dynamic measure of hyperuniformity and nonhyperuniformity in heterogeneous media via the diffusion spreadability},\ }\href {https://doi.org/https://doi.org/10.1103/PhysRevApplied.17.034022} {\bibfield  {journal} {\bibinfo  {journal} {Phys. Rev. Appl.}\ }\textbf {\bibinfo {volume} {17}},\ \bibinfo {pages} {034022} (\bibinfo {year} {2022})}\BibitemShut {NoStop}%
\bibitem [{\citenamefont {Pauling}(1949)}]{pauling1949resonating}%
  \BibitemOpen
  \bibfield  {author} {\bibinfo {author} {\bibfnamefont {L.~C.}\ \bibnamefont {Pauling}},\ }\bibfield  {title} {\bibinfo {title} {A resonating-valence-bond theory of metals and intermetallic compounds},\ }\href {https://doi.org/10.1098/rspa.1949.0032} {\bibfield  {journal} {\bibinfo  {journal} {Proc. R. Soc. A}\ }\textbf {\bibinfo {volume} {196}},\ \bibinfo {pages} {343} (\bibinfo {year} {1949})}\BibitemShut {NoStop}%
\bibitem [{\citenamefont {Anderson}(1973)}]{anderson1973resonating}%
  \BibitemOpen
  \bibfield  {author} {\bibinfo {author} {\bibfnamefont {P.~W.}\ \bibnamefont {Anderson}},\ }\bibfield  {title} {\bibinfo {title} {{Resonating valence bonds: A new kind of insulator?}},\ }\href {https://doi.org/10.1016/0025-5408(73)90167-0} {\bibfield  {journal} {\bibinfo  {journal} {Mat. Res. Bull.}\ }\textbf {\bibinfo {volume} {8}},\ \bibinfo {pages} {153} (\bibinfo {year} {1973})}\BibitemShut {NoStop}%
\bibitem [{\citenamefont {Baskaran}\ and\ \citenamefont {Anderson}(1988)}]{baskaran1988gauge}%
  \BibitemOpen
  \bibfield  {author} {\bibinfo {author} {\bibfnamefont {G.}~\bibnamefont {Baskaran}}\ and\ \bibinfo {author} {\bibfnamefont {P.~W.}\ \bibnamefont {Anderson}},\ }\bibfield  {title} {\bibinfo {title} {{Gauge theory of high-temperature superconductors and strongly correlated Fermi systems}},\ }\href {https://doi.org/10.1103/PhysRevB.37.580} {\bibfield  {journal} {\bibinfo  {journal} {Phys. Rev. B}\ }\textbf {\bibinfo {volume} {37}},\ \bibinfo {pages} {580} (\bibinfo {year} {1988})}\BibitemShut {NoStop}%
\bibitem [{\citenamefont {Rokhsar}\ and\ \citenamefont {Kivelson}(1988)}]{rokhsar1988superconductivity}%
  \BibitemOpen
  \bibfield  {author} {\bibinfo {author} {\bibfnamefont {D.~S.}\ \bibnamefont {Rokhsar}}\ and\ \bibinfo {author} {\bibfnamefont {S.~A.}\ \bibnamefont {Kivelson}},\ }\bibfield  {title} {\bibinfo {title} {{Superconductivity and the Quantum Hard-Core Dimer Gas}},\ }\href {https://doi.org/10.1103/PhysRevLett.61.2376} {\bibfield  {journal} {\bibinfo  {journal} {Phys. Rev. Lett.}\ }\textbf {\bibinfo {volume} {61}},\ \bibinfo {pages} {2376} (\bibinfo {year} {1988})}\BibitemShut {NoStop}%
\bibitem [{\citenamefont {Selem}\ \emph {et~al.}(2013)\citenamefont {Selem}, \citenamefont {Herdman},\ and\ \citenamefont {Whaley}}]{Se13}%
  \BibitemOpen
  \bibfield  {author} {\bibinfo {author} {\bibfnamefont {A.}~\bibnamefont {Selem}}, \bibinfo {author} {\bibfnamefont {C.~M.}\ \bibnamefont {Herdman}},\ and\ \bibinfo {author} {\bibfnamefont {K.~B.}\ \bibnamefont {Whaley}},\ }\bibfield  {title} {\bibinfo {title} {Entanglement entropy at generalized Rokhsar-Kivelson points of quantum dimer models},\ }\href {https://doi.org/https://doi.org/10.1103/PhysRevB.87.125105} {\bibfield  {journal} {\bibinfo  {journal} {Phys. Rev. B}\ }\textbf {\bibinfo {volume} {87}},\ \bibinfo {pages} {125105} (\bibinfo {year} {2013})}\BibitemShut {NoStop}%
\bibitem [{\citenamefont {Sutherland}(1988{\natexlab{a}})}]{Su88}%
  \BibitemOpen
  \bibfield  {author} {\bibinfo {author} {\bibfnamefont {B.}~\bibnamefont {Sutherland}},\ }\bibfield  {title} {\bibinfo {title} {Systems with resonating-valence-bond ground states: Correlations and excitations},\ }\href {https://doi.org/https://doi.org/10.1103/PhysRevB.37.3786} {\bibfield  {journal} {\bibinfo  {journal} {Phys. Rev. B}\ }\textbf {\bibinfo {volume} {37}},\ \bibinfo {pages} {3786} (\bibinfo {year} {1988}{\natexlab{a}})}\BibitemShut {NoStop}%
\bibitem [{\citenamefont {Read}\ and\ \citenamefont {Chakraborty}(1989)}]{read1989statistics}%
  \BibitemOpen
  \bibfield  {author} {\bibinfo {author} {\bibfnamefont {N.}~\bibnamefont {Read}}\ and\ \bibinfo {author} {\bibfnamefont {B.}~\bibnamefont {Chakraborty}},\ }\bibfield  {title} {\bibinfo {title} {Statistics of the excitations of the resonating-valence-bond state},\ }\href {https://doi.org/10.1103/PhysRevB.40.7133} {\bibfield  {journal} {\bibinfo  {journal} {Phys. Rev. B}\ }\textbf {\bibinfo {volume} {40}},\ \bibinfo {pages} {7133} (\bibinfo {year} {1989})}\BibitemShut {NoStop}%
\bibitem [{\citenamefont {Senthil}\ and\ \citenamefont {Fisher}(2000)}]{SenthilFisher}%
  \BibitemOpen
  \bibfield  {author} {\bibinfo {author} {\bibfnamefont {T.}~\bibnamefont {Senthil}}\ and\ \bibinfo {author} {\bibfnamefont {M.~P.~A.}\ \bibnamefont {Fisher}},\ }\bibfield  {title} {\bibinfo {title} {${Z}_{2}$ gauge theory of electron fractionalization in strongly correlated systems},\ }\href {https://doi.org/10.1103/PhysRevB.62.7850} {\bibfield  {journal} {\bibinfo  {journal} {Phys. Rev. B}\ }\textbf {\bibinfo {volume} {62}},\ \bibinfo {pages} {7850} (\bibinfo {year} {2000})}\BibitemShut {NoStop}%
\bibitem [{\citenamefont {Sutherland}(1988{\natexlab{b}})}]{sutherland1988systems}%
  \BibitemOpen
  \bibfield  {author} {\bibinfo {author} {\bibfnamefont {B.}~\bibnamefont {Sutherland}},\ }\bibfield  {title} {\bibinfo {title} {{Systems with resonating-valence-bond ground states: Correlations and excitations}},\ }\href {https://doi.org/10.1103/PhysRevB.37.3786} {\bibfield  {journal} {\bibinfo  {journal} {Phys. Rev. B}\ }\textbf {\bibinfo {volume} {37}},\ \bibinfo {pages} {3786} (\bibinfo {year} {1988}{\natexlab{b}})}\BibitemShut {NoStop}%
\bibitem [{\citenamefont {Moessner}\ and\ \citenamefont {Sondhi}(2001{\natexlab{b}})}]{moessner2001resonating}%
  \BibitemOpen
  \bibfield  {author} {\bibinfo {author} {\bibfnamefont {R.}~\bibnamefont {Moessner}}\ and\ \bibinfo {author} {\bibfnamefont {S.~L.}\ \bibnamefont {Sondhi}},\ }\bibfield  {title} {\bibinfo {title} {{Resonating Valence Bond Phase in the Triangular Lattice Quantum Dimer Model}},\ }\href {https://doi.org/10.1103/PhysRevLett.86.1881} {\bibfield  {journal} {\bibinfo  {journal} {Phys. Rev. Lett.}\ }\textbf {\bibinfo {volume} {86}},\ \bibinfo {pages} {1881} (\bibinfo {year} {2001}{\natexlab{b}})}\BibitemShut {NoStop}%
\bibitem [{\citenamefont {{Moessner}}\ \emph {et~al.}(2002)\citenamefont {{Moessner}}, \citenamefont {{Sondhi}},\ and\ \citenamefont {{Fradkin}}}]{MSF02}%
  \BibitemOpen
  \bibfield  {author} {\bibinfo {author} {\bibfnamefont {R.}~\bibnamefont {{Moessner}}}, \bibinfo {author} {\bibfnamefont {S.~L.}\ \bibnamefont {{Sondhi}}},\ and\ \bibinfo {author} {\bibfnamefont {E.}~\bibnamefont {{Fradkin}}},\ }\bibfield  {title} {\bibinfo {title} {{Short-ranged resonating valence bond physics, quantum dimer models, and Ising gauge theories}},\ }\href {https://doi.org/10.1103/PhysRevB.65.024504} {\bibfield  {journal} {\bibinfo  {journal} {Phys. Rev. B}\ }\textbf {\bibinfo {volume} {65}},\ \bibinfo {eid} {024504} (\bibinfo {year} {2002})}\BibitemShut {NoStop}%
\bibitem [{\citenamefont {Samajdar}\ \emph {et~al.}(2023)\citenamefont {Samajdar}, \citenamefont {Joshi}, \citenamefont {Teng},\ and\ \citenamefont {Sachdev}}]{PhysRevLett.130.043601}%
  \BibitemOpen
  \bibfield  {author} {\bibinfo {author} {\bibfnamefont {R.}~\bibnamefont {Samajdar}}, \bibinfo {author} {\bibfnamefont {D.~G.}\ \bibnamefont {Joshi}}, \bibinfo {author} {\bibfnamefont {Y.}~\bibnamefont {Teng}},\ and\ \bibinfo {author} {\bibfnamefont {S.}~\bibnamefont {Sachdev}},\ }\bibfield  {title} {\bibinfo {title} {{Emergent ${\mathbb{Z}}_{2}$ Gauge Theories and Topological Excitations in Rydberg Atom Arrays}},\ }\href {https://doi.org/10.1103/PhysRevLett.130.043601} {\bibfield  {journal} {\bibinfo  {journal} {Phys. Rev. Lett.}\ }\textbf {\bibinfo {volume} {130}},\ \bibinfo {pages} {043601} (\bibinfo {year} {2023})}\BibitemShut {NoStop}%
\bibitem [{\citenamefont {White}(1992)}]{white1992density}%
  \BibitemOpen
  \bibfield  {author} {\bibinfo {author} {\bibfnamefont {S.~R.}\ \bibnamefont {White}},\ }\bibfield  {title} {\bibinfo {title} {Density matrix formulation for quantum renormalization groups},\ }\href {https://doi.org/10.1103/PhysRevLett.69.2863} {\bibfield  {journal} {\bibinfo  {journal} {Phys. Rev. Lett.}\ }\textbf {\bibinfo {volume} {69}},\ \bibinfo {pages} {2863} (\bibinfo {year} {1992})}\BibitemShut {NoStop}%
\bibitem [{\citenamefont {White}(1993)}]{white1993density}%
  \BibitemOpen
  \bibfield  {author} {\bibinfo {author} {\bibfnamefont {S.~R.}\ \bibnamefont {White}},\ }\bibfield  {title} {\bibinfo {title} {Density-matrix algorithms for quantum renormalization groups},\ }\href {https://doi.org/10.1103/PhysRevB.48.10345} {\bibfield  {journal} {\bibinfo  {journal} {Phys. Rev. B}\ }\textbf {\bibinfo {volume} {48}},\ \bibinfo {pages} {10345} (\bibinfo {year} {1993})}\BibitemShut {NoStop}%
\bibitem [{\citenamefont {Schollw{\"o}ck}(2005)}]{schollwock2005density}%
  \BibitemOpen
  \bibfield  {author} {\bibinfo {author} {\bibfnamefont {U.}~\bibnamefont {Schollw{\"o}ck}},\ }\bibfield  {title} {\bibinfo {title} {The density-matrix renormalization group},\ }\href {https://doi.org/10.1103/RevModPhys.77.259} {\bibfield  {journal} {\bibinfo  {journal} {Rev. Mod. Phys.}\ }\textbf {\bibinfo {volume} {77}},\ \bibinfo {pages} {259} (\bibinfo {year} {2005})}\BibitemShut {NoStop}%
\bibitem [{\citenamefont {Schollw{\"o}ck}(2011)}]{schollwock2011density}%
  \BibitemOpen
  \bibfield  {author} {\bibinfo {author} {\bibfnamefont {U.}~\bibnamefont {Schollw{\"o}ck}},\ }\bibfield  {title} {\bibinfo {title} {The density-matrix renormalization group: a short introduction},\ }\href {https://doi.org/10.1098/rsta.2010.0382} {\bibfield  {journal} {\bibinfo  {journal} {Phil. Trans. R. Soc. A}\ }\textbf {\bibinfo {volume} {369}},\ \bibinfo {pages} {2643} (\bibinfo {year} {2011})}\BibitemShut {NoStop}%
\bibitem [{\citenamefont {Nikolic}\ and\ \citenamefont {Senthil}(2003)}]{nikolic2003physics}%
  \BibitemOpen
  \bibfield  {author} {\bibinfo {author} {\bibfnamefont {P.}~\bibnamefont {Nikolic}}\ and\ \bibinfo {author} {\bibfnamefont {T.}~\bibnamefont {Senthil}},\ }\bibfield  {title} {\bibinfo {title} {{Physics of low-energy singlet states of the Kagome lattice quantum Heisenberg antiferromagnet}},\ }\href {https://doi.org/10.1103/PhysRevB.68.214415} {\bibfield  {journal} {\bibinfo  {journal} {Phys. Rev. B}\ }\textbf {\bibinfo {volume} {68}},\ \bibinfo {pages} {214415} (\bibinfo {year} {2003})}\BibitemShut {NoStop}%
\bibitem [{\citenamefont {Singh}\ and\ \citenamefont {Huse}(2007)}]{singh2007ground}%
  \BibitemOpen
  \bibfield  {author} {\bibinfo {author} {\bibfnamefont {R.~R.~P.}\ \bibnamefont {Singh}}\ and\ \bibinfo {author} {\bibfnamefont {D.~A.}\ \bibnamefont {Huse}},\ }\bibfield  {title} {\bibinfo {title} {{Ground state of the spin-1/2 kagome-lattice Heisenberg antiferromagnet}},\ }\href {https://doi.org/10.1103/PhysRevB.76.180407} {\bibfield  {journal} {\bibinfo  {journal} {Phys. Rev. B}\ }\textbf {\bibinfo {volume} {76}},\ \bibinfo {pages} {180407} (\bibinfo {year} {2007})}\BibitemShut {NoStop}%
\bibitem [{\citenamefont {Poilblanc}\ and\ \citenamefont {Misguich}(2011)}]{poilblanc2011competing}%
  \BibitemOpen
  \bibfield  {author} {\bibinfo {author} {\bibfnamefont {D.}~\bibnamefont {Poilblanc}}\ and\ \bibinfo {author} {\bibfnamefont {G.}~\bibnamefont {Misguich}},\ }\bibfield  {title} {\bibinfo {title} {Competing valence bond crystals in the kagome quantum dimer model},\ }\href {https://doi.org/10.1103/PhysRevB.84.214401} {\bibfield  {journal} {\bibinfo  {journal} {Phys. Rev. B}\ }\textbf {\bibinfo {volume} {84}},\ \bibinfo {pages} {214401} (\bibinfo {year} {2011})}\BibitemShut {NoStop}%
\bibitem [{\citenamefont {Kitaev}\ and\ \citenamefont {Preskill}(2006)}]{PhysRevLett.96.110404}%
  \BibitemOpen
  \bibfield  {author} {\bibinfo {author} {\bibfnamefont {A.}~\bibnamefont {Kitaev}}\ and\ \bibinfo {author} {\bibfnamefont {J.}~\bibnamefont {Preskill}},\ }\bibfield  {title} {\bibinfo {title} {{Topological Entanglement Entropy}},\ }\href {https://doi.org/10.1103/PhysRevLett.96.110404} {\bibfield  {journal} {\bibinfo  {journal} {Phys. Rev. Lett.}\ }\textbf {\bibinfo {volume} {96}},\ \bibinfo {pages} {110404} (\bibinfo {year} {2006})}\BibitemShut {NoStop}%
\bibitem [{\citenamefont {Jiang}\ \emph {et~al.}(2012)\citenamefont {Jiang}, \citenamefont {Wang},\ and\ \citenamefont {Balents}}]{jiang2012identifying}%
  \BibitemOpen
  \bibfield  {author} {\bibinfo {author} {\bibfnamefont {H.-C.}\ \bibnamefont {Jiang}}, \bibinfo {author} {\bibfnamefont {Z.}~\bibnamefont {Wang}},\ and\ \bibinfo {author} {\bibfnamefont {L.}~\bibnamefont {Balents}},\ }\bibfield  {title} {\bibinfo {title} {Identifying topological order by entanglement entropy},\ }\href {https://doi.org/10.1038/nphys2465} {\bibfield  {journal} {\bibinfo  {journal} {Nature Physics}\ }\textbf {\bibinfo {volume} {8}},\ \bibinfo {pages} {902} (\bibinfo {year} {2012})}\BibitemShut {NoStop}%
\bibitem [{\citenamefont {Torquato}(2021)}]{To21b}%
  \BibitemOpen
  \bibfield  {author} {\bibinfo {author} {\bibfnamefont {S.}~\bibnamefont {Torquato}},\ }\bibfield  {title} {\bibinfo {title} {Structural characterization of many-particle systems on approach to hyperuniform states},\ }\href {https://doi.org/https://doi.org/10.1103/PhysRevE.103.052126} {\bibfield  {journal} {\bibinfo  {journal} {Phys. Rev. E}\ }\textbf {\bibinfo {volume} {103}},\ \bibinfo {pages} {052126} (\bibinfo {year} {2021})}\BibitemShut {NoStop}%
\bibitem [{\citenamefont {Kim}\ and\ \citenamefont {Torquato}(2017)}]{Ki17}%
  \BibitemOpen
  \bibfield  {author} {\bibinfo {author} {\bibfnamefont {J.}~\bibnamefont {Kim}}\ and\ \bibinfo {author} {\bibfnamefont {S.}~\bibnamefont {Torquato}},\ }\bibfield  {title} {\bibinfo {title} {Effect of window shape on the detection of hyperuniformity via the local number variance},\ }\href {https://doi.org/https://doi.org/10.1088/1742-5468/aa4f9d} {\bibfield  {journal} {\bibinfo  {journal} {J. Stat. Mech. Theory Exp.}\ }\textbf {\bibinfo {volume} {2017}},\ \bibinfo {pages} {013402} (\bibinfo {year} {2017})}\BibitemShut {NoStop}%
\bibitem [{\citenamefont {Batten}\ \emph {et~al.}(2008)\citenamefont {Batten}, \citenamefont {Stillinger},\ and\ \citenamefont {Torquato}}]{Ba08}%
  \BibitemOpen
  \bibfield  {author} {\bibinfo {author} {\bibfnamefont {R.~D.}\ \bibnamefont {Batten}}, \bibinfo {author} {\bibfnamefont {F.~H.}\ \bibnamefont {Stillinger}},\ and\ \bibinfo {author} {\bibfnamefont {S.}~\bibnamefont {Torquato}},\ }\bibfield  {title} {\bibinfo {title} {Classical disordered ground states: Super-ideal gases and stealth and equi-luminous materials},\ }\href {https://doi.org/https://doi.org/10.1063/1.2961314} {\bibfield  {journal} {\bibinfo  {journal} {J. Appl. Phys.}\ }\textbf {\bibinfo {volume} {104}},\ \bibinfo {pages} {033504} (\bibinfo {year} {2008})}\BibitemShut {NoStop}%
\bibitem [{\citenamefont {Torquato}\ \emph {et~al.}(2015)\citenamefont {Torquato}, \citenamefont {Zhang},\ and\ \citenamefont {Stillinger}}]{To15}%
  \BibitemOpen
  \bibfield  {author} {\bibinfo {author} {\bibfnamefont {S.}~\bibnamefont {Torquato}}, \bibinfo {author} {\bibfnamefont {G.}~\bibnamefont {Zhang}},\ and\ \bibinfo {author} {\bibfnamefont {F.~H.}\ \bibnamefont {Stillinger}},\ }\bibfield  {title} {\bibinfo {title} {Ensemble theory for stealthy hyperuniform disordered ground states},\ }\href {https://doi.org/https://doi.org/10.1103/PhysRevX.5.021020} {\bibfield  {journal} {\bibinfo  {journal} {Phys. Rev. X}\ }\textbf {\bibinfo {volume} {5}},\ \bibinfo {pages} {021020} (\bibinfo {year} {2015})}\BibitemShut {NoStop}%
\bibitem [{\citenamefont {Hitin-Bialus}\ \emph {et~al.}(2024)\citenamefont {Hitin-Bialus}, \citenamefont {Maher}, \citenamefont {Steinhardt},\ and\ \citenamefont {Torquato}}]{Hi24}%
  \BibitemOpen
  \bibfield  {author} {\bibinfo {author} {\bibfnamefont {A.}~\bibnamefont {Hitin-Bialus}}, \bibinfo {author} {\bibfnamefont {C.~E.}\ \bibnamefont {Maher}}, \bibinfo {author} {\bibfnamefont {P.~J.}\ \bibnamefont {Steinhardt}},\ and\ \bibinfo {author} {\bibfnamefont {S.}~\bibnamefont {Torquato}},\ }\bibfield  {title} {\bibinfo {title} {Hyperuniformity classes of quasiperiodic tilings via diffusion spreadability},\ }\href {https://doi.org/https://doi.org/10.1103/PhysRevE.109.064108} {\bibfield  {journal} {\bibinfo  {journal} {Phys. Rev. E}\ }\textbf {\bibinfo {volume} {109}},\ \bibinfo {pages} {064108} (\bibinfo {year} {2024})}\BibitemShut {NoStop}%
\bibitem [{\citenamefont {Giudice}\ \emph {et~al.}(2022)\citenamefont {Giudice}, \citenamefont {Surace}, \citenamefont {Pichler},\ and\ \citenamefont {Giudici}}]{PhysRevB.106.195155}%
  \BibitemOpen
  \bibfield  {author} {\bibinfo {author} {\bibfnamefont {G.}~\bibnamefont {Giudice}}, \bibinfo {author} {\bibfnamefont {F.~M.}\ \bibnamefont {Surace}}, \bibinfo {author} {\bibfnamefont {H.}~\bibnamefont {Pichler}},\ and\ \bibinfo {author} {\bibfnamefont {G.}~\bibnamefont {Giudici}},\ }\bibfield  {title} {\bibinfo {title} {{Trimer states with ${\mathbb{Z}}_{3}$ topological order in Rydberg atom arrays}},\ }\href {https://doi.org/10.1103/PhysRevB.106.195155} {\bibfield  {journal} {\bibinfo  {journal} {Phys. Rev. B}\ }\textbf {\bibinfo {volume} {106}},\ \bibinfo {pages} {195155} (\bibinfo {year} {2022})}\BibitemShut {NoStop}%
\bibitem [{\citenamefont {Kornja{\v{c}}a}\ \emph {et~al.}(2023)\citenamefont {Kornja{\v{c}}a}, \citenamefont {Samajdar}, \citenamefont {Macr{\`\i}}, \citenamefont {Gemelke}, \citenamefont {Wang},\ and\ \citenamefont {Liu}}]{kornjavca2022trimer}%
  \BibitemOpen
  \bibfield  {author} {\bibinfo {author} {\bibfnamefont {M.}~\bibnamefont {Kornja{\v{c}}a}}, \bibinfo {author} {\bibfnamefont {R.}~\bibnamefont {Samajdar}}, \bibinfo {author} {\bibfnamefont {T.}~\bibnamefont {Macr{\`\i}}}, \bibinfo {author} {\bibfnamefont {N.}~\bibnamefont {Gemelke}}, \bibinfo {author} {\bibfnamefont {S.-T.}\ \bibnamefont {Wang}},\ and\ \bibinfo {author} {\bibfnamefont {F.}~\bibnamefont {Liu}},\ }\bibfield  {title} {\bibinfo {title} {Trimer quantum spin liquid in a honeycomb array of rydberg atoms},\ }\href {https://doi.org/10.1038/s42005-023-01470-z} {\bibfield  {journal} {\bibinfo  {journal} {Commun. Phys.}\ }\textbf {\bibinfo {volume} {6}},\ \bibinfo {pages} {358} (\bibinfo {year} {2023})}\BibitemShut {NoStop}%
\bibitem [{\citenamefont {Snyder}\ \emph {et~al.}(2001)\citenamefont {Snyder}, \citenamefont {Slusky}, \citenamefont {Cava},\ and\ \citenamefont {Schiffer}}]{Sn01}%
  \BibitemOpen
  \bibfield  {author} {\bibinfo {author} {\bibfnamefont {J.}~\bibnamefont {Snyder}}, \bibinfo {author} {\bibfnamefont {J.~S.}\ \bibnamefont {Slusky}}, \bibinfo {author} {\bibfnamefont {R.~J.}\ \bibnamefont {Cava}},\ and\ \bibinfo {author} {\bibfnamefont {P.}~\bibnamefont {Schiffer}},\ }\bibfield  {title} {\bibinfo {title} {How `spin ice' freezes},\ }\href {https://doi.org/https://doi.org/10.1038/35092516} {\bibfield  {journal} {\bibinfo  {journal} {Nature}\ }\textbf {\bibinfo {volume} {413}},\ \bibinfo {pages} {48} (\bibinfo {year} {2001})}\BibitemShut {NoStop}%
\bibitem [{\citenamefont {Scheie}\ \emph {et~al.}(2024{\natexlab{a}})\citenamefont {Scheie}, \citenamefont {Lee}, \citenamefont {Wang}, \citenamefont {Laurell}, \citenamefont {Choi}, \citenamefont {Pajerowski}, \citenamefont {Zhang}, \citenamefont {Ma}, \citenamefont {Zhou}, \citenamefont {Lee} \emph {et~al.}}]{Sc24b}%
  \BibitemOpen
  \bibfield  {author} {\bibinfo {author} {\bibfnamefont {A.~O.}\ \bibnamefont {Scheie}}, \bibinfo {author} {\bibfnamefont {M.}~\bibnamefont {Lee}}, \bibinfo {author} {\bibfnamefont {K.}~\bibnamefont {Wang}}, \bibinfo {author} {\bibfnamefont {P.}~\bibnamefont {Laurell}}, \bibinfo {author} {\bibfnamefont {E.~S.}\ \bibnamefont {Choi}}, \bibinfo {author} {\bibfnamefont {D.}~\bibnamefont {Pajerowski}}, \bibinfo {author} {\bibfnamefont {Q.}~\bibnamefont {Zhang}}, \bibinfo {author} {\bibfnamefont {J.}~\bibnamefont {Ma}}, \bibinfo {author} {\bibfnamefont {H.~D.}\ \bibnamefont {Zhou}}, \bibinfo {author} {\bibfnamefont {S.}~\bibnamefont {Lee}}, \emph {et~al.},\ }\bibfield  {title} {\bibinfo {title} {Spectrum and low-energy gap in triangular quantum spin liquid NaYbSe$_2$},\ }\href {https://arxiv.org/abs/2406.17773} {\bibfield  {journal} {\bibinfo  {journal} {arXiv preprint arXiv:2406.17773}\ } (\bibinfo {year} {2024}{\natexlab{a}})}\BibitemShut {NoStop}%
\bibitem [{\citenamefont {Scheie}\ \emph {et~al.}(2024{\natexlab{b}})\citenamefont {Scheie}, \citenamefont {Ghioldi}, \citenamefont {Xing}, \citenamefont {Paddison}, \citenamefont {Sherman}, \citenamefont {Dupont}, \citenamefont {Sanjeewa}, \citenamefont {Lee}, \citenamefont {Woods}, \citenamefont {Abernathy} \emph {et~al.}}]{Sc24}%
  \BibitemOpen
  \bibfield  {author} {\bibinfo {author} {\bibfnamefont {A.~O.}\ \bibnamefont {Scheie}}, \bibinfo {author} {\bibfnamefont {E.~A.}\ \bibnamefont {Ghioldi}}, \bibinfo {author} {\bibfnamefont {J.}~\bibnamefont {Xing}}, \bibinfo {author} {\bibfnamefont {J.~A.~M.}\ \bibnamefont {Paddison}}, \bibinfo {author} {\bibfnamefont {N.~E.}\ \bibnamefont {Sherman}}, \bibinfo {author} {\bibfnamefont {M.}~\bibnamefont {Dupont}}, \bibinfo {author} {\bibfnamefont {L.~D.}\ \bibnamefont {Sanjeewa}}, \bibinfo {author} {\bibfnamefont {S.}~\bibnamefont {Lee}}, \bibinfo {author} {\bibfnamefont {A.~J.}\ \bibnamefont {Woods}}, \bibinfo {author} {\bibfnamefont {D.}~\bibnamefont {Abernathy}}, \emph {et~al.},\ }\bibfield  {title} {\bibinfo {title} {Proximate spin liquid and fractionalization in the triangular antiferromagnet KYbSe$_2$},\ }\href {https://doi.org/https://doi.org/10.1038/s41567-023-02259-1} {\bibfield  {journal} {\bibinfo  {journal} {Nat. Phys.}\ }\textbf {\bibinfo {volume} {20}},\ \bibinfo {pages} {74} (\bibinfo {year}
  {2024}{\natexlab{b}})}\BibitemShut {NoStop}%
\bibitem [{\citenamefont {Yeong}\ and\ \citenamefont {Torquato}(1998{\natexlab{a}})}]{Ye98a}%
  \BibitemOpen
  \bibfield  {author} {\bibinfo {author} {\bibfnamefont {C.~L.~Y.}\ \bibnamefont {Yeong}}\ and\ \bibinfo {author} {\bibfnamefont {S.}~\bibnamefont {Torquato}},\ }\bibfield  {title} {\bibinfo {title} {Reconstructing random media},\ }\href {https://doi.org/https://doi.org/10.1103/PhysRevE.57.495} {\bibfield  {journal} {\bibinfo  {journal} {Phys. Rev. E}\ }\textbf {\bibinfo {volume} {57}},\ \bibinfo {pages} {495} (\bibinfo {year} {1998}{\natexlab{a}})}\BibitemShut {NoStop}%
\bibitem [{\citenamefont {Yeong}\ and\ \citenamefont {Torquato}(1998{\natexlab{b}})}]{Ye98b}%
  \BibitemOpen
  \bibfield  {author} {\bibinfo {author} {\bibfnamefont {C.~L.~Y.}\ \bibnamefont {Yeong}}\ and\ \bibinfo {author} {\bibfnamefont {S.}~\bibnamefont {Torquato}},\ }\bibfield  {title} {\bibinfo {title} {{Reconstructing random media. II. Three-dimensional media from two-dimensional cuts}},\ }\href {https://doi.org/https://doi.org/10.1103/PhysRevE.58.224} {\bibfield  {journal} {\bibinfo  {journal} {Phys. Rev. E}\ }\textbf {\bibinfo {volume} {58}},\ \bibinfo {pages} {224} (\bibinfo {year} {1998}{\natexlab{b}})}\BibitemShut {NoStop}%
\bibitem [{\citenamefont {Fishman}\ \emph {et~al.}(2022)\citenamefont {Fishman}, \citenamefont {White},\ and\ \citenamefont {Stoudenmire}}]{ITensor}%
  \BibitemOpen
  \bibfield  {author} {\bibinfo {author} {\bibfnamefont {M.}~\bibnamefont {Fishman}}, \bibinfo {author} {\bibfnamefont {S.~R.}\ \bibnamefont {White}},\ and\ \bibinfo {author} {\bibfnamefont {E.~M.}\ \bibnamefont {Stoudenmire}},\ }\bibfield  {title} {\bibinfo {title} {{The ITensor Software Library for Tensor Network Calculations}},\ }\href {https://doi.org/10.21468/SciPostPhysCodeb.4} {\bibfield  {journal} {\bibinfo  {journal} {SciPost Phys. Codebases}\ ,\ \bibinfo {pages} {4}} (\bibinfo {year} {2022})}\BibitemShut {NoStop}%
\end{thebibliography}
%

\end{document}